\documentclass[10pt,  letterpaper]{IEEEtran}
\usepackage{todonotes}
\usepackage[nocompress]{cite}
\usepackage{url}
\usepackage{enumitem}
\usepackage{diagbox} 
\usepackage{epstopdf}
\PrependGraphicsExtensions{.tif, .tiff}
\usepackage{graphicx}
\usepackage{caption}
\usepackage{amsmath}
\usepackage{tabu}
\usepackage{multirow}
\usepackage{fontawesome}
\usepackage{adjustbox}
\usepackage{float}
\usepackage{mathptmx}
\usepackage{amsfonts}
\usepackage{algpseudocode}
\usepackage{soul}
\usepackage[bookmarks=false]{hyperref}
\usepackage{footnote}
\usepackage{lipsum}
\usepackage[ruled,vlined, linesnumbered]{algorithm2e}
\usepackage{subcaption}
\usepackage{listings}
\usepackage{booktabs}
\usepackage{tikz}
\usetikzlibrary{positioning}
\soulregister\cite7
\soulregister\ref7
\soulregister\pageref7
\soulregister\secref7
\soulregister\tabref7
\definecolor{ao}{rgb}{0.0, 0.5, 0.0}
\definecolor{db}{rgb}{0.2, 0.2, 0.6}
\definecolor{cadmiumgreen}{rgb}{0.0, 0.42, 0.24}

\newcommand{\tabref}[1]{Table~\ref{#1}}

\newcommand{\secref}[1]{Section~\ref{#1}}

\SetAlFnt{\footnotesize}

\newcommand{\shrinkspace}{\vspace{-5mm}}

\begin{document}
% \title{Training Heterogeneous Client Models using Knowledge Distillation in Serverless Federated Learning}
% \author{\IEEEauthorblockN{Mohamed Elzohairy\IEEEauthorrefmark{1}, Mohak Chadha\IEEEauthorrefmark{1}, Anshul Jindal\IEEEauthorrefmark{1}, Andreas Grafberger\IEEEauthorrefmark{1}, Jianfeng Gu\IEEEauthorrefmark{1}, \\ Michael Gerndt\IEEEauthorrefmark{1}, Osama Abboud\IEEEauthorrefmark{2}\\}
% \IEEEauthorblockA{\IEEEauthorrefmark{1}Chair of Computer Architecture and Parallel Systems, Technische Universit{\"a}t M{\"u}nchen \\
% Garching (near Munich), Germany \\}
% \IEEEauthorblockA{\IEEEauthorrefmark{2}Huawei Technologies, Munich, Germany \\} 
% Email: \{mohamed.elzohairy, mohak.chadha, anshul.jindal, andreas.grafberger, jianfeng.gu\}@tum.de, \\gerndt@in.tum.de, osama.abboud@huawei.com}

\title{Apodotiko: Enabling Efficient Serverless Federated Learning in Heterogeneous Environments}
\author{\IEEEauthorblockN{Mohak Chadha\IEEEauthorrefmark{1},   Alexander Jensen\IEEEauthorrefmark{1}, Jianfeng Gu\IEEEauthorrefmark{1}, Osama Abboud\IEEEauthorrefmark{2}, Michael Gerndt\IEEEauthorrefmark{1}\\}
\IEEEauthorblockA{\IEEEauthorrefmark{1}Chair of Computer Architecture and Parallel Systems, Technische Universit{\"a}t M{\"u}nchen \\
Garching (near Munich), Germany \\}
\IEEEauthorblockA{\IEEEauthorrefmark{2}Huawei Technologies, Munich, Germany \\} 
Email: \{mohak.chadha, alexander.jensen, jianfeng.gu\}@tum.de, osama.abboud@huawei.com, gerndt@in.tum.de}

% make the title area
\maketitle
%Uncomment before final submission
% \IEEEpubidadjcol
\pagenumbering{gobble}
%\copyrightnotice
%  \copyrightnotice

% As a general rule, do not put math, special symbols or citations
% in the abstract

\begin{abstract}

% Federated Learning (FL) is a machine learning paradigm that enables the training of a shared global model across distributed clients  While most prior work on designing systems for FL has focused on using stateful always running components, recent work has shown that components in an FL system can greatly benefit from the usage of serverless computing and Function-as-a-Service technologies. To this end, distributed training of models with serverless FL systems can be more resource-efficient and cheaper than conventional FL systems. 
Federated Learning (FL) is an emerging machine learning paradigm that enables the collaborative training of a shared global model across distributed clients while keeping the data decentralized. Recent works on designing systems for efficient FL have shown that utilizing serverless computing technologies, particularly Function-as-a-Service (FaaS) for FL, can enhance resource efficiency, reduce training costs, and alleviate the complex infrastructure management burden on data holders. However, current serverless FL systems still suffer from the presence of \emph{stragglers}, i.e., slow clients that impede the collaborative training process. While strategies aimed at mitigating stragglers in these systems have been proposed, they overlook the diverse hardware resource configurations among FL clients. To this end, we present \texttt{Apodotiko}, a novel \emph{asynchronous} training strategy designed for serverless FL. Our strategy incorporates a scoring mechanism that evaluates each client's hardware capacity and dataset size to intelligently prioritize and select clients for each training round, thereby minimizing the effects of stragglers on system performance. We comprehensively evaluate \texttt{Apodotiko} across diverse datasets, considering a mix of CPU and GPU clients, and compare its performance against five other FL training strategies. Results from our experiments demonstrate that \texttt{Apodotiko} outperforms other FL training strategies, achieving an average speedup of $2.75$x and a maximum speedup of $7.03$x. Furthermore, our strategy significantly reduces cold starts by a factor of four on average, demonstrating suitability in serverless environments.

\end{abstract}


% \IEEEpeerreviewmaketitle
% \thispagestyle{empty}

\shrinkspace

\section{Introduction}
\label{sec:intro}

Increasing concerns about data privacy and recent legislations such as the Consumer Privacy Bill of Rights in the U.S.~\cite{Privacy_and_big_data} prevent the training of ML models using the traditional centralized learning approach.  With the goal of not exposing raw data as in centralized learning~\cite{lecun2015deep}, an emerging distributed training paradigm called Federated Learning (FL)~\cite{mcmahan2017communication} has gained significant popularity in various application domains, such as medical care~\cite{rieke2020future} and mobile services~\cite{huba2022papaya}.

FL enables the collaborative training of a shared global ML model across remote devices or \texttt{clients} while keeping the training data decentralized. The traditional FL training process~\cite{mcmahan2017communication} is \textit{synchronous} and occurs in multiple rounds. A main component called the \texttt{central} \texttt{server} organizes the training process and decides which clients contribute in a new round. During each round, clients improve the shared global model by optimizing it on their local datasets and sending back only the updated model parameters to the central server. Following this, the local model updates from all participating clients are collected and aggregated to form the updated consensus model. Recent works on designing systems for efficient FL have shown that both components in an FL system, i.e., the \texttt{clients} and the \texttt{central} \texttt{server}, can immensely benefit from an emerging cloud computing paradigm called \textit{serverless computing}~\cite{serverlessfl, fedless, elzohairy2022fedlesscan, jayaram2022lambda, jitfl, jayaramadaptive, flox}.

Function-as-a-Service (FaaS) is the computational concept of serverless computing and has gained significant popularity and widespread adoption in various application domains such as machine learning~\cite{fastgshare, chadha2024training}, edge computing~\cite{fado}, heterogeneous computing~\cite{fncapacitor, jindal2021function, courier, postericdcs}, and scientific computing~\cite{chadha2021architecture, demystifying}. In FaaS,  developers implement fine-grained pieces of code called \textit{functions} that are packaged independently in containers and uploaded onto a FaaS platform. These functions are \textit{ephemeral}, \textit{event-driven}, and \textit{stateless}. Several open-source and commercial FaaS platforms, such as OpenFaaS~\cite{openfaas} and Google Cloud Functions (GCF)~\cite{gcloud-functions-2}, are currently available. Clients in serverless FL are independent functions deployed onto a FaaS platform and capable of performing their model updates.

The FaaS computing model offers several advantages, such as no infrastructure management, automatic scaling to zero when resources are unused, and an attractive fine-grained \textit{pay-per-use} billing policy~\cite{castro2019rise}.
Incorporating FaaS functions as \texttt{clients}  in FL systems can improve resource efficiency and reduce training costs~\cite{fedless, flox}. In addition, utilizing FaaS technologies for the aggregation process in the FL \texttt{server} can enhance aggregation performance, scalability, and resource efficiency~\cite{fedless, jayaram2022lambda, jayaramadaptive, jitfl}.

% \begin{figure*}[t]
% \begin{subfigure}{0.33\textwidth}
%     \centering
%         \includegraphics[width=\columnwidth]{figures/motivation_comparison_acc.pdf}
%         \caption{Comparing trained model accuracy.}
%         \label{fig:comparingacc}
% \end{subfigure}
% \begin{subfigure}{0.33\textwidth}
%     \centering
%         \includegraphics[width=\columnwidth]{figures/motivation_comparison_round.pdf}
%         \caption{Comparing FL round durations.}
%         \label{fig:comprounduration}
% \end{subfigure}
% % \hspace{-5mm}
% \begin{subfigure}{0.34\textwidth}
%     \centering
%         \includegraphics[width=0.32\columnwidth]{figures/motivation_var_cpu_same.pdf}
%         \includegraphics[width=0.32\columnwidth]{figures/motivation_var_cpu_mix.pdf}
%         \includegraphics[width=0.32\columnwidth]{figures/motivation_var_cpu_gpu_mix.pdf}
%         \vspace{0.01mm}
%         \caption{Comparing Selection Bias.}
%         \label{fig:compbias}
% \end{subfigure}
% \shrinkspace
% \caption{Comparing \texttt{FedAvg}~\cite{mcmahan2017communication} and \texttt{FedLesScan}~\cite{elzohairy2022fedlesscan} across various client-hardware resource configurations using \emph{FedLess}~\cite{fedless}. The results are obtained using the Shakespeare dataset~\cite{caldas2018leaf} with 100 clients deployed on \texttt{OpenFaaS}~\cite{openfaas}.}
% \shrinkspace
% \label{fig:motivation}
% \end{figure*}

Large-scale practical FL systems encounter different fundamental client-level challenges that limit collaborative training. These include computational heterogeneity and statistical data heterogeneity. FL clients in the wild~\cite{huba2022papaya} can vary from small edge devices to high-performant GPU-enabled systems with varying memory, compute, and storage capacities. In addition, clients in practical FL systems have \textit{unbalanced non-IID} data distributions, i.e., the private data samples held by individual clients exhibit variations in their statistical properties, such as feature distributions, class imbalances, or data biases~\cite{hsieh2020non}. These two challenges often result in the presence of \texttt{stragglers}, i.e., slower clients within the FL system. Stragglers tend to increase the duration and costs of the FL training process while diminishing the accuracy of the trained global model~\cite{elzohairy2022fedlesscan, chai2020tifl, pisces}.

% To mitigate stragglers in FL, several strategies have been proposed in the literature. These strategies can be classified into three categories: (i) \emph{synchronous}~\cite{mcmahan2017communication, fedprox, scaffold}, (ii) \emph{semi-asynchronous}~\cite{wu2020safa}, and (iii) \emph{asynchronous}~\cite{fedbuff, pisces}. However, most of these techniques fail to consider the unique characteristics of serverless environments, such as function cold starts, performance variations, and the transient stateless nature of function instances. To this end, \texttt{FedLesScan}~\cite{elzohairy2022fedlesscan} represents the current state-of-the-art for straggler mitigation in serverless FL. It is a \emph{semi-asynchronous} training strategy that dynamically adapts to the behavior of clients to minimize the impact of stragglers on the FL system. It consists of two key components: a clustering-based client selection algorithm and a staleness-aware aggregation scheme. The former is responsible for selecting a subset of clients for training based on their previous training round durations, while the latter accounts for delayed client round updates.

To mitigate stragglers in FL, several strategies have been proposed in the literature~\cite{mcmahan2017communication, fedprox, scaffold, wu2020safa, fedbuff, pisces}. However, most of these techniques fail to consider the unique characteristics of serverless environments, such as function cold starts, performance variations, and the transient stateless nature of function instances. To this end, \texttt{FedLesScan}~\cite{elzohairy2022fedlesscan} represents the current state-of-the-art for straggler mitigation in serverless FL. It is a \emph{semi-asynchronous} training strategy that dynamically adapts to the behavior of clients to minimize the impact of stragglers on the FL system. It consists of two key components: a clustering-based client selection algorithm and a staleness-aware aggregation scheme. The former is responsible for selecting a subset of clients for training based on their previous training round durations, while the latter accounts for delayed client round updates.

% It incorporates an intelligent clustering-based client selection algorithm that selects a subset of clients for training based on their previous training round durations. In addition, it includes a staleness-aware aggregation scheme to account for delayed round updates.

% However, this approach fails to accommodate the increasing heterogeneity in client hardware and private data size, leading to sub-optimal cluster sizes for sampling clients in each round. As a result, stragglers are inadvertently included in the training round, limiting the benefits of \texttt{FedLesScan}. 

\begin{figure}[t]
\begin{subfigure}{0.48\textwidth}
    \centering
        \includegraphics[width=\columnwidth]{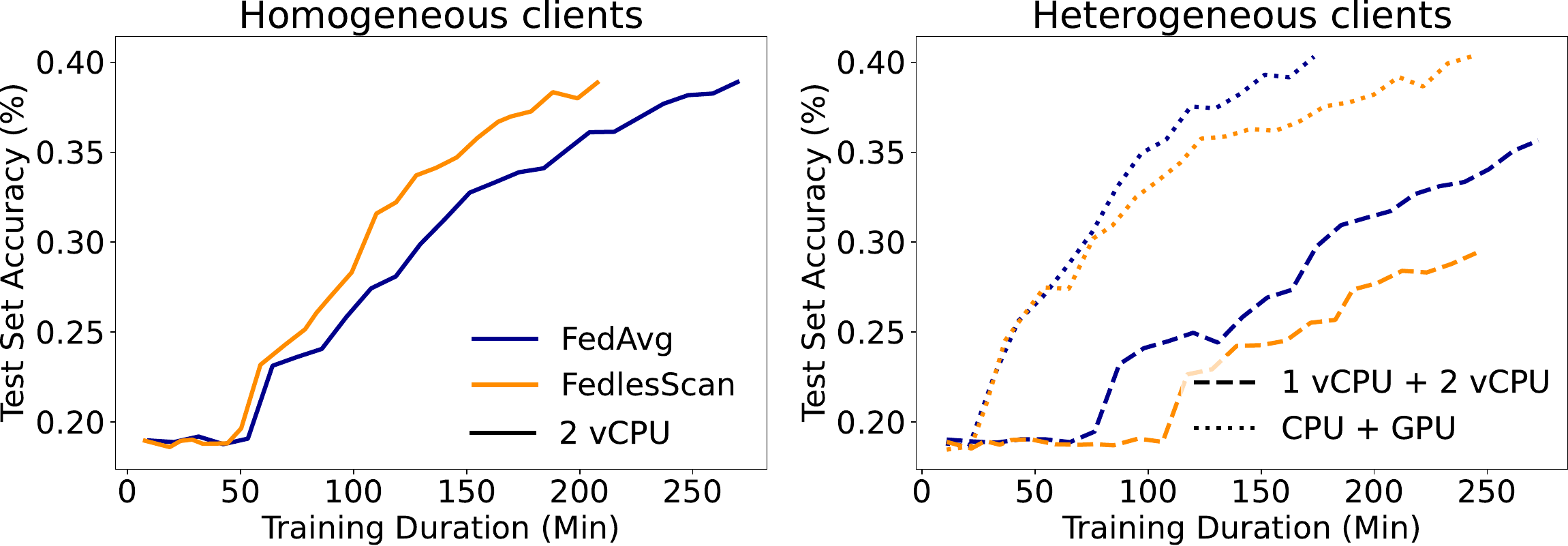}
        \caption{Comparing trained model accuracy.}
        \label{fig:comparingacc}
\end{subfigure}
\begin{subfigure}{0.48\textwidth}
    \centering
        \includegraphics[width=\columnwidth]{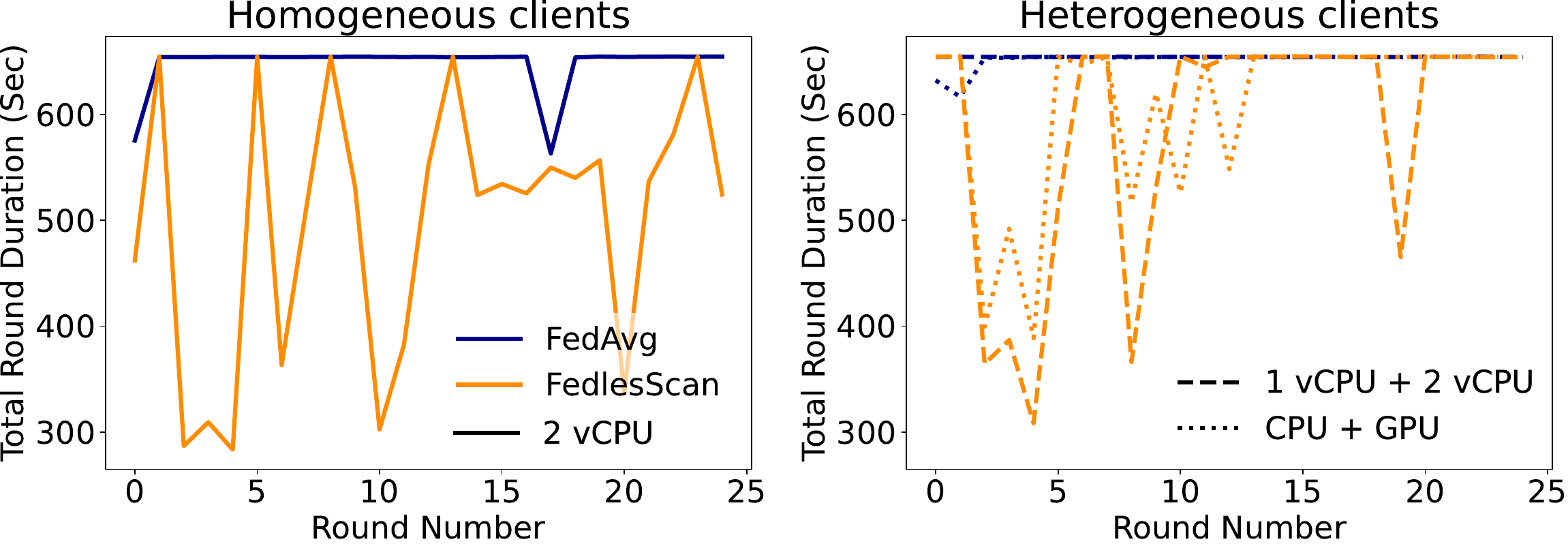}
        \caption{Comparing FL round durations.}
        \label{fig:comprounduration}
\end{subfigure}
% \hspace{-5mm}
% \shrinkspace
\vspace{-3mm}
\caption{Comparing \texttt{FedAvg}~\cite{mcmahan2017communication} and \texttt{FedLesScan}~\cite{elzohairy2022fedlesscan} across various client-hardware resource configurations using \emph{FedLess}~\cite{fedless}. The results are obtained using the non-IID data partitions of the \emph{Shakespeare} dataset~\cite{caldas2018leaf} with 100 clients deployed on \texttt{OpenFaaS}~\cite{openfaas}.}
\shrinkspace
\vspace{-2mm}
\label{fig:motivation}
\end{figure}

Figure~\ref{fig:motivation} compares the performance of \texttt{FedLesScan} with \texttt{FedAvg} across three different client hardware resource distribution scenarios. In the first scenario, all $100$ clients have the same hardware resource configuration of $2$vCPUs. In the second scenario, we configured $60$ clients with $1$vCPU and $40$ clients with $2$vCPUs, while in the third scenario, we had a mix of 50 clients with $1$vCPU, 30 clients with $2$vCPUs, and 20 clients running on GPUs. In each round, $50$ clients are selected for training. With homogeneous clients, we observe that \texttt{FedLesScan} requires $30$\% less time than \texttt{FedAvg} to achieve an accuracy of $40$\% for the global model as shown in Figure~\ref{fig:comparingacc}. However, with heterogeneous clients, \texttt{FedlesScan} struggles and lags behind \texttt{FedAvg} in performance. For the second scenario, \texttt{FedlesScan} requires $40$\% more training time than \texttt{FedAvg} to reach an accuracy of $30$\%, while for the third scenario \texttt{FedAvg} is $43$\% faster. To provide deeper insights, Figure~\ref{fig:comprounduration} offers a detailed breakdown of the training duration for individual rounds. In the homogeneous scenario, \texttt{FedlesScan} effectively clusters clients with adequate sizes, enabling efficient client selection for each training round. This leads to sporadic peaks in round durations, occurring when stragglers are included. However, in the heterogeneous scenarios, more rounds reach their maximum duration, resembling the round times observed with \texttt{FedAvg}. This is because the clustering method adopted by \texttt{FedlesScan} fails to provide enough clients per cluster for selection in each round, resulting in the inclusion of stragglers as replacements. The presence of stragglers during training rounds causes the FL server to time out, resulting in relatively consistent round durations. Our experiments and observations demonstrate that \texttt{FedlesScan} fails to accommodate the increasing heterogeneity in FL client hardware under non-IID data distributions, leading to decreased global model accuracy and increased training times. To address these limitations and advance the state-of-the-art in serverless FL, our key contributions are:
\begin{itemize}
    \item We present \texttt{Apodotiko}\footnote{\url{https://github.com/Serverless-Federated-Learning/FedLess}}, a novel \emph{asynchronous} scoring-based strategy that enables efficient serverless FL across clients with varying hardware resource configurations and data distributions. 
    \item We implement \texttt{Apodotiko} by extending a popular open-source serverless FL framework called \emph{FedLess}~\cite{fedless, osstalk}. This enables greater adoption and accessibility of our strategy in the community.
    \item We demonstrate with extensive experiments the performance of our strategy against five other popular FL training approaches across multiple datasets using various metrics, including accuracy, selection bias, and costs.
\end{itemize}

The rest of the paper is structured as follows. In \S\ref{sec:relatedwork}, we describe the previous approaches related to our work. \S\ref{sec:apodotiko} describes our strategy in detail. In \S\ref{sec:expresults}, we present our experimental results. Finally, \S\ref{sec:concfuture} concludes the paper and presents an outlook.

\vspace{-4mm}

\section{Related Work}
\label{sec:relatedwork}

\subsection{Serverless Federated Learning}
\label{sec:serverlessfl}

Using serverless computing technologies, particularly FaaS, for designing efficient systems for FL is a relatively new research direction. Existing works in this domain can be categorized into two groups: (i) systems that employ serverless functions exclusively in the \texttt{central server} ~\cite{jayaram2022lambda, jitfl, jayaramadaptive} and (ii) systems that leverage serverless functions in both entities of an FL system~\cite{flox, serverlessfl, fedless}. In~\cite{jayaram2022lambda}, Jayaram et al. propose \(\lambda\)-FL, a serverless aggregation strategy for FL to improve fault tolerance and reduce resource wastage. The authors use serverless functions as aggregators to optimize the aggregation of model parameters in conventional \texttt{FedAvg}~\cite{mcmahan2017communication} over several steps. They implement their prototype using message queues, \texttt{Kafka}, and use \texttt{Ray}~\cite{ray} as the serverless platform. In~\cite{jitfl} and~\cite{jayaramadaptive}, the authors extend their previous strategy to enable adaptive and just-in-time aggregation of client model updates using serverless functions. In the second group, \emph{FedKeeper}~\cite{serverlessfl} was the first serverless FL system that enabled the training of Deep Neural Network (DNN) models using FL for clients distributed across a combination of heterogeneous FaaS platforms. However, it lacked crucial features required for practical FL systems, such as security and support for large DNN models. To address these drawbacks, \emph{FedLess}~\cite{fedless} was introduced as an evolution of \emph{FedKeeper} with multiple new enhancements. These include: (i) support for multiple open-source and commercial FaaS platforms, (ii) authentication/authorization of client functions using AWS Cognito, (iii) training of arbitrary homogeneous DNN models using the \texttt{Tensorflow} library, (iv) the privacy-preserving FL training of models using Differential Privacy~\cite{Mothukuri2021}, and (v) the aggregation of model updates using serverless functions. A more recent work by Kotsehub et al.~\cite{flox} introduces \emph{Flox}, a system built on the \texttt{funcX}~\cite{chard2020funcx} serverless platform. \emph{Flox} aims to separate FL model training/inference from infrastructure management, providing users with a convenient way to deploy FL models on heterogeneous distributed compute endpoints. However, its tight integration with \texttt{funcX} restricts its compatibility with other open-source or commercial FaaS platforms, limiting its applicability and generality. 

% As a result, in this paper, we use \emph{FedLess} as the serverless FL framework for implementing our strategy. 

% In addition, \emph{FedLess} incorporates several optimizations for serverless environments, such as global namespace caching, running average model aggregation, and federated evaluation. 

\vspace{-4mm}

\subsection{Stragglers in Federated Learning}
\label{sec:stragglersfl}
Addressing the impact of slow clients during training in FL is an active research area. Towards this, several \emph{synchronous}~\cite{li2020federated, wang2020tackling, scaffold, chai2020tifl, cox2022aergia, Oort}, \emph{semi-asynchronous}~\cite{wu2020safa, elzohairy2022fedlesscan}, and \emph{asynchronous}~\cite{chai2021fedat, fedasync, fedbuff, pisces} strategies have been proposed in the literature. 

% In \emph{synchronous} FL, the central server waits for all participating clients to complete their training and send their updates before aggregating these updates and proceeding to the next round. In contrast, \emph{asynchronous} FL enables clients to train and communicate with the central server independently, without waiting for other clients. 

% \emph{Synchronous} FL training strategies can often lead to increased training times and costs, while \emph{asynchronous} approaches have demonstrated convergence rates, albeit at higher communication costs.

\begin{table}[t]
\begin{adjustbox}{width=\columnwidth,  center}
\begin{tabular}{|c|c|c|c|c|c|c|}
\hline
\multicolumn{1}{|l|}{\diagbox{Strategy}{Attribute}} & Type & \multicolumn{1}{c|}{\begin{tabular}[c]{@{}c@{}}FaaS \\ Support\end{tabular}} & \multicolumn{1}{c|}{\begin{tabular}[c]{@{}c@{}}Asynchronous \\ Aggregation\end{tabular}} & \multicolumn{1}{c|}{\begin{tabular}[c]{@{}c@{}}Performance-based \\ Selection\end{tabular}} & \multicolumn{1}{c|}{\begin{tabular}[c]{@{}c@{}}Client Efficiency \\ Scoring\end{tabular}} & \multicolumn{1}{c|}{\begin{tabular}[c]{@{}c@{}}Adaptive\\ Penalty\end{tabular}} \\ \hline
\textbf{FedProx}~\cite{li2020federated} & Synchronous      &    \faThumbsDown                                                                           &            \faThumbsDown                                                                               &        \faThumbsDown                                                                                      &          \faThumbsDown                                                                                  &       \faThumbsDown                                                                           \\ \hline
\textbf{FedNova}~\cite{wang2020tackling}   & Synchronous     &          \faThumbsDown                                                                      &            \faThumbsDown                                                                                &                    \faThumbsDown                                                                           &           \faThumbsDown                                                                                  &          \faThumbsDown                                                                         \\ \hline
\textbf{SCAFFOLD}~\cite{scaffold}      & Synchronous   &                   \faThumbsDown                                                           &                                  \faThumbsDown                                                          &                               \faThumbsDown                                                                &                      \faThumbsDown                                                                       &                 \faThumbsDown                                                                  \\ \hline
\textbf{TiFL}~\cite{chai2020tifl}          & Synchronous &                      \faThumbsDown                                                          &                             \faThumbsDown                                                               &                        \faThumbsOUp                                                                    &                     \faThumbsDown                                                                        &               \faThumbsOUp                                                                \\ \hline
\textbf{Aergia}~\cite{cox2022aergia}      & Synchronous   &           \faThumbsDown                                                                   &                   \faThumbsDown                                                                       &              \faThumbsDown                                                                               &       \faThumbsDown                                                                                    &    \faThumbsDown                                                                             \\ \hline
\textbf{Oort}~\cite{Oort}          &     Synchronous  &                            \faThumbsDown                                             &                                           \faThumbsDown                                               &                                   \faThumbsOUp                                                            &                             \faThumbsDown                                                              &                         \faThumbsDown                                                        \\ \hline
\textbf{SAFA}~\cite{wu2020safa}          &   Semi-asynchronous  &           \faThumbsDown                                                                    &                          \faThumbsDown                                                                  &                      \faThumbsDown                                                                         &              \faThumbsDown                                                                               &          \faThumbsDown                                                                         \\ \hline
\textbf{FedAT}~\cite{chai2021fedat}    &   Semi-asynchronous      &                       \faThumbsDown                                                         &                                    \faThumbsDown                                                        &                               \faThumbsOUp                                                                &                        \faThumbsDown                                                                     &                  \faThumbsOUp                                                                 \\ \hline
\textbf{FedAsync}~\cite{fedasync}      &    Asynchronous      &               \faThumbsDown                                                             &                          \faThumbsOUp                                                              &                          \faThumbsDown                                                                   &                     \faThumbsDown                                                                      &                 \faThumbsDown                                                                \\ \hline
\textbf{FedBuff}~\cite{fedbuff}       &        Asynchronous      &            \faThumbsDown                                                           &                               \faThumbsOUp                                                           &                            \faThumbsDown                                                                 &                       \faThumbsDown                                                                    &                   \faThumbsDown                                                              \\ \hline
\textbf{Pisces}~\cite{pisces}          &       Asynchronous      &                  \faThumbsDown                                                      &                                    \faThumbsOUp                                                    &                                  \faThumbsOUp                                                           &                              \faThumbsDown                                                             &                         \faThumbsOUp                                                     \\ \hline
\textbf{FedlesScan}~\cite{elzohairy2022fedlesscan}     &   Semi-asynchronous      &      \faThumbsOUp                                                                        &                     \faThumbsDown                                                                     &                    \faThumbsOUp                                                                           &               \faThumbsDown                                                                            &           \faThumbsOUp                                                                     \\ \hline
\textbf{Apodotiko} (This work)     &       Asynchronous      &                   \faThumbsOUp                                                    &                                      \faThumbsOUp                                                    &                                          \faThumbsOUp                                                   &                                      \faThumbsOUp                                                     &                                     \faThumbsOUp                                            \\ \hline
\end{tabular}
\end{adjustbox}
\caption{Comparing strategies for straggler mitigation in FL. \faThumbsOUp  Supported. \faThumbsDown No support.}
\shrinkspace
% \vspace{-2mm}
\label{tab:strategies_comparison}
\end{table}

\texttt{FedProx}~\cite{li2020federated}, a popular strategy, builds on \texttt{FedAvg} with two essential modifications. First, it introduces a specialized loss function at the client level to regulate fluctuations in local updates, enhancing the model's stability across varied data distributions. Second, it enables clients to adjust their workload based on hardware and network constraints, ensuring adaptability by varying the number of local updates performed. Similar to \texttt{FedProx}, \texttt{FedNova}~\cite{wang2020tackling} tackles statistical heterogeneity in FL by merging the concepts of \texttt{FedAvg} with momentum-based optimization. It introduces a momentum term to stabilize convergence and alleviate the impact of noisy local updates. To accommodate varying local updates for clients, it introduces a new weighting scheme that normalizes client local updates with the number of local steps. However, \texttt{FedNova} is tailored only for \texttt{SGD}, limiting its broader applicability. In~\cite{scaffold}, Karimireddy et al. propose \texttt{SCAFFOLD}, an FL strategy designed to address the challenges of varied local client data distributions and biased updates. The authors employ \emph{control variates}, a technique from convex optimization, to reduce stochastic gradient variance. This minimizes local update variability, stabilizing the aggregation process. In addition, it enables identifying and eliminating client-specific biases pre-aggregation, enhancing global model accuracy and stability. 
Unlike \texttt{FedProx}, \texttt{SCAFFOLD} doesn't accommodate varying local progress and mandates uniform local update counts across clients. Moreover, it relies on full client participation for peak performance, diminishing its effectiveness with reduced client sampling ratios per round, as shown in~\cite{li2022federated}. \texttt{TiFL}~\cite{chai2020tifl} is a tier-based system designed to address heterogeneity challenges in FL (\S\ref{sec:intro}). It organizes clients into tiers, selecting same-tier clients per round to mitigate the impact of stragglers. In addition, it incorporates an adaptive tiering mechanism that dynamically adjusts tiers based on observed training performance. However, the authors limit their experiments to only CPU-based clients without exploring extreme heterogeneous environments with a mix of both CPU and GPU-based clients. In~\cite{Oort}, Lai et al. propose \texttt{Oort}~\cite{Oort} a strategy that aims to optimize FL training by prioritizing clients that offer the most valuable contributions to model accuracy.  It assesses clients based on their utility in improving model accuracy and their ability to train efficiently while preserving privacy. To select high-utility clients, \texttt{Oort} employs an online exploration-exploitation strategy that dynamically adapts the selection process to account for outliers and achieve a balance between statistical and system efficiency. Unlike other \emph{synchronous} FL strategies, \texttt{Aergia}~\cite{cox2022aergia} freezes the computationally intensive part of a slow client's model and offloads it to a faster client that trains it using its own dataset. It leverages the spare computational capacity from robust clients 
and achieves high accuracy in relatively low training times by effectively matching clients' performance profiles and data similarity.

\texttt{SAFA}~\cite{wu2020safa} is a \emph{semi-asynchronous} training strategy that targets improved round efficiency and convergence, especially in scenarios with frequent client dropouts. It introduces innovative client selection and global aggregation methods, including a caching mechanism to prevent wasted client contributions. \texttt{FedAT}~\cite{chai2021fedat} uses a tiering mechanism that partitions clients into logical tiers based on their response latencies. Faster tiers update the model synchronously, while slower tiers send updates asynchronously. It employs a weighted aggregation approach to avoid bias toward faster tiers and uses compression to reduce communication costs. In~\cite{fedasync}, Xie et al. propose \texttt{FedAsync}~\cite{fedasync}, an \emph{asynchronous} FL strategy that utilizes a parameter server architecture to synchronize and coordinate client invocations. It employs a scheduler thread to trigger client training and an updater thread for aggregating client updates into the global model. To address scalability concerns in practical FL systems, \texttt{FedBuff}~\cite{fedbuff} introduces buffered asynchronous aggregations. In this strategy, clients train and communicate asynchronously with the server, storing their updates in a buffer until a server update triggers aggregation once a specific number of client updates are collected. In~\cite{pisces}, Jiang et al. propose \texttt{Pisces}~\cite{pisces}, an \emph{asynchronous} FL strategy that utilizes guided participant selection and adaptive aggregation pacing to mitigate slow clients. It merges techniques from \texttt{FedBuff} and \texttt{Oort} to enhance performance, focusing on prioritizing participants with high data quality as \texttt{Oort}.  This enables more efficient utilization of clients compared to \texttt{FedBuff}.

Table~\ref{tab:strategies_comparison} provides a comprehensive comparison between \texttt{Apodotiko} and the different strategies for straggler mitigation in FL. We differentiate these strategies based on five attributes: support for FaaS environments, asynchronous aggregation, performance-based selection, client-efficiency scoring, and adaptive penalty. FaaS support indicates compatibility with serverless environments. Asynchronous aggregation reflects the flexibility to separate client updates from training. Performance-based selection involves choosing clients based on their training duration. Client efficiency scoring accounts for hardware diversity during selection, while adaptive penalty reflects adjustments in client selection based on performance and availability over time. While \texttt{FedProx}, \texttt{FedNova}, \texttt{SCAFFOLD}, and \texttt{Aergia} primarily focus on optimizing the local training process and aggregation methods, they do not incorporate intelligent client selection to optimize round performance as done in \texttt{Apodotiko}. \texttt{SAFA} tracks the status of clients' local models to ensure their synchronization with the global model but tends to overutilize clients and lacks suitability for FaaS environments. \texttt{TiFL}, \texttt{FedAT}, and \texttt{FedlesScan} group clients based on training duration into clusters, but they overlook the hardware and data heterogeneity during the client selection process and lack support for asynchronous aggregation. Although \texttt{Oort} considers data size and training duration in client selection, it overlooks the correlation between these factors and diverse hardware configurations, enforcing a strict penalty on slower clients. \texttt{FedAsync} and \texttt{FedBuff} focus on optimizing FL with asynchronous aggregation but adopt random client selection. In contrast, \texttt{Pisces} combines the methods from \texttt{Oort} and \texttt{FedBuff}, refining the scoring approach, yet it still overlooks clients' efficiency during scoring (\S\ref{sec:scorebased}) and selection. \texttt{Apodotiko} overcomes these limitations by incorporating comprehensive scoring metrics that account for both hardware and data heterogeneity, ensuring intelligent client selection and efficient round performance.

\vspace{-4mm}

\section{Apodotiko}
\label{sec:apodotiko}
In this section, we describe our strategy for \emph{asynchronous} serverless FL in detail. First, we provide an overview of \texttt{Apodotiko}, followed by our methodology for \emph{asynchronous aggregation} of model updates. After this, we present our novel client scoring strategy, followed by our probabilistic client selection technique.

\vspace{-3mm}

\subsection{Overview}
\label{sec:overview}
The \emph{FedLess}~\cite{fedless} framework includes a central component known as the controller, which oversees and orchestrates the entire FL training process. It incorporates a Strategy Manager subcomponent~\cite{ elzohairy2022fedlesscan} that is responsible for controlling the behavior of the selected strategy. This involves managing client selection and choosing the aggregation technique used. To implement \texttt{Apodotiko}, we extend the Strategy Manager subcomponent in \emph{FedLess}. In addition, to enable asynchronous aggregation (\S\ref{sec:asyncaggregation}) and collect different client attributes required for our scoring strategy (\S\ref{sec:scorebased}), we modify the controller and client routines as shown in Algorithm~\ref{alg:routine}. At the start of each FL training round, the controller runs the \texttt{Train\_Global\_Model} routine, while the selected clients execute the \texttt{Client\_Update} routine to locally train the global model.

\setlength{\textfloatsep}{0pt}
\begin{algorithm}[t]
  \SetAlgoLined
  \SetKwFunction{train}{Client\_Update}
  \SetKwFunction{selectclients}{Select\_Clients}
   \SetKwFunction{sleep}{Sleep}
  \SetKwFunction{fitRound}{Train\_Global\_Model}
  \SetKwFunction{startTime}{Start\_Timer}
  \SetKwFunction{stopTime}{Stop\_Timer}
  \SetKwProg{Pn}{Function}{:}{\KwRet}
  \DontPrintSemicolon
  \textbf{Fedless Controller:}
  \;
  \Pn{\fitRound{$clients$, $round$}}
  {
    $client\_selection$ = \selectclients{$clients$, $clientsPerRound$}\;
    Invoke selected clients\;
    \For{each $client$ in $client\_selection$}{
      Save invocation record to database ($client$, $round$)\;
      Set $client$ invocation status to $runnning$ \tcp*[l]{Busy client}\label{alg:line:busy}
    }
    \While{$\#results \ge (clientsPerRound * concurrencyRatio)$}{\label{lst:line:asyn}
      \sleep{$0.1$}
    }
    Aggregate Model\;
  }
  \textbf{Fedless Client:}
  \;
  \Pn{\train{$hyperParameters$, $round$}}
  {
    Load model and dataset.\;
    \startTime{}\label{alg:line:routine_start_time}\;
    Train model\;
    \stopTime{}\label{alg:line:routine_stop_time}\;
    Save updated model to database. \;
    Add measured time to invocation record in database.\;
    Update $invocation$ status to $complete$. \tcp*[l]{Available client}\label{alg:line:available}
  }
  \caption{Modifed \emph{FedLess}~\cite{fedless} controller and client routines.}
  % \shrinkspace
  \label{alg:routine}
\end{algorithm}

Initially, the controller selects the required number of clients using our client selection strategy (\S\ref{sec:selectingclients}) and invokes them (Lines 3-4). Following this, we store the information regarding the invoked clients, such as the current round number, in the \emph{FedLess} database (Line 6). Moreover, we mark the currently running clients as \emph{busy} (Line 7). This is required to prevent selecting already running clients in the next FL training round. After this, the controller periodically checks for the availability of results from a fraction of clients in the database and then triggers the global model aggregation function (\S\ref{sec:asyncaggregation}, \S\ref{sec:serverlessfl}) (Lines 9-12). On the client side, the selected clients retrieve the most recent global model along with their local datasets and perform local model updates for a specified number of configured epochs (Lines 16-19). Following this, they store the updated model and the measured training time in the database (Lines 20-21). Finally, we mark the finished clients as \emph{complete} in the database, making them available for selection in the next FL training rounds (Line 22).

\begin{footnotesize}
\begin{equation}
  w_{t+1} \longleftarrow   \sum_{i=1}^{N} \frac{t_i}{T} \times \frac{n_{i}}{n} w^i_{t_i}
  \label{eqn:staleness_old}
\end{equation}
% \shrinkspace
\end{footnotesize}

\begin{footnotesize}
\begin{equation}
  w_{t+1} \longleftarrow    \sum_{i=1}^{N} \frac{1}{(T - t_i + 1)^{0.5}} \times \frac{n_{i}}{n} w^i_{t_i}
  \label{eqn:staleness_new}
\end{equation}
\shrinkspace
\end{footnotesize}

\subsection{Asynchronous Aggregation}
\label{sec:asyncaggregation}
\begin{figure}[t]
\begin{subfigure}{0.23\textwidth}
    \centering
        \includegraphics[width=0.7\columnwidth]{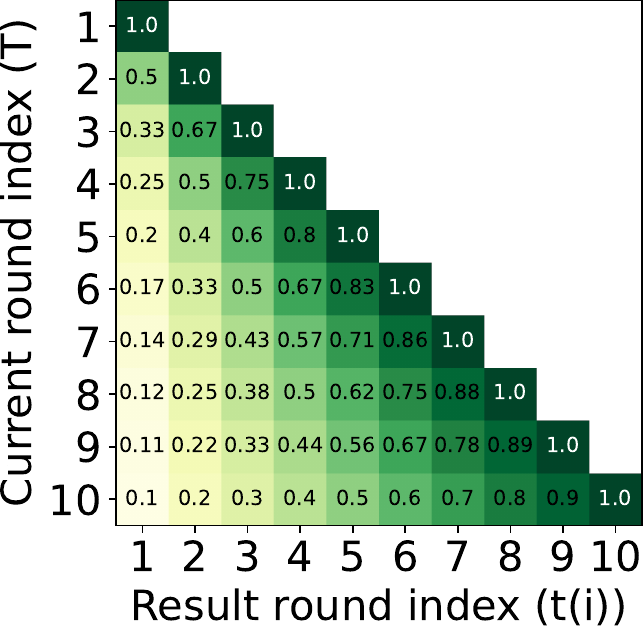}
        \caption{\texttt{FedLesScan}.}
        \label{fig:equationprev}
\end{subfigure}
\begin{subfigure}{0.23\textwidth}
    \centering
        \includegraphics[width=0.7\columnwidth]{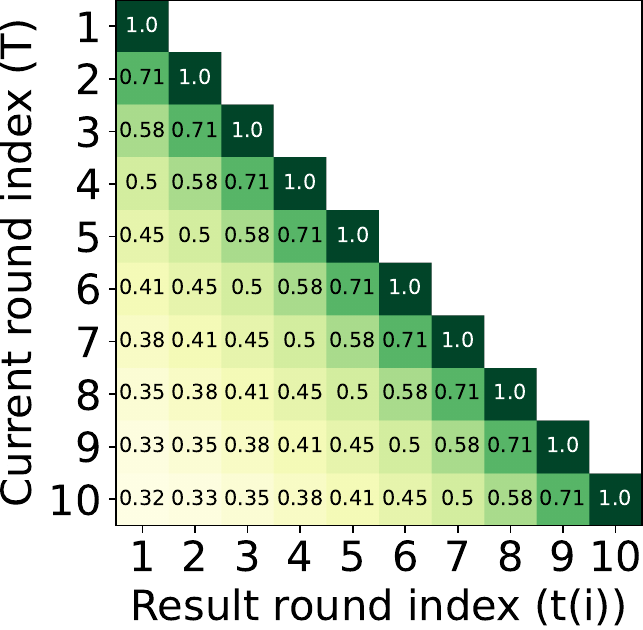}
        \caption{\texttt{Apodotiko}.}
        \label{fig:equationnew}
\end{subfigure}
% \hspace{-5mm}
\vspace{-1mm}
\caption{Comparing weighting functions for aggregating stale client model updates.}
% \shrinkspace
% \vspace{-0.5mm}
%Better caption
\end{figure}

In \emph{synchronous} FL, implementing per-round timeouts is a common strategy to prevent exceedingly long round times~\cite{fedless, scaffold, mcmahan2017communication}. This timeout mechanism ensures that the central server does not wait indefinitely for all clients to send their updates before initiating the global model aggregation. However, slow clients might push their local model updates to the parameter server after the completion of an FL training round. These updates, often discarded, can potentially contain valuable information that can improve the performance of the global model. To reduce training times and improve convergence rates, we extend the \emph{FedLess}~\cite{fedless} controller to aggregate client model updates asynchronously without waiting for all current client model updates to be available in the database. Towards this, the modified controller in \texttt{Apodotiko} only waits for a fraction of client updates, referred to as the \emph{concurrencyRatio} ((0,1]), before invoking the aggregator function. For instance, with $100$ clients per round and a ratio of $0.6$, the controller only waits for updates from $60$ clients. These client updates can be from the current or previous training rounds. Aggregating model updates from previous rounds can lead to reduced convergence rate and higher variance in the global model~\cite{fedasync}. Moreover, the older the update, the higher the risk to the quality of the global model. To mitigate this, most \emph{asynchronous} strategies utilize a staleness weighting function to dampen updates from previous rounds during aggregation. This weighting function should assign a weight of 1 to the current round's results and show a monotonically decreasing pattern with increasing round numbers. With \texttt{Apodotiko}, we experimented with different staleness functions shown in Equations~\ref{eqn:staleness_old} and~\ref{eqn:staleness_new}. The former is used by \texttt{FedLesScan}~\cite{elzohairy2022fedlesscan}, while the latter is adopted from~\cite{fedasync}. In these equations, $w^i_{t_i}$ represents the local model of the client \textit{i} at round $t_i$, while $w_{t+1}$ is the global model after aggregation at round $T$. Furthermore, $n_i$ represents the cardinality of the dataset at client $i$ while $n$ is the total cardinality of the aggregated clients. With Equation~\ref{eqn:staleness_old}, we observe that the weight of one round of late updates gradually increases as the round number increases, as shown in Figure~\ref{fig:equationprev}. Moreover, in Figure~\ref{fig:equationprev}, the weight values derived from Equation~\ref{eqn:staleness_old} exhibit inconsistency for results with identical staleness levels, contrasting Figure~\ref{fig:equationnew} with Eq.~\ref{eqn:staleness_new}, where the weight values maintain consistency along the diagonal axis. As a result, we use Eq.~\ref{eqn:staleness_new} with our strategy for aggregating model updates. In our experiments, we only considered client updates from a maximum of five previous rounds but observed that most delayed updates arrived within two rounds. Although conceptually similar, our \emph{asynchronous aggregation} mechanism in \texttt{Apodotiko} differs from the buffering technique used in \texttt{FedBuff}~\cite{fedbuff}. In \texttt{FedBuff}, client model updates are stored in-memory, whereas in \texttt{Apodotiko}, we utilize an external database for this purpose. 
Furthermore, our approach to aggregating stale model updates differs from the method outlined in~\cite{fedbuff}.
% Moreover, we utilize a different strategy for aggregating stale model updates as compared to~\cite{fedbuff}.

\begin{figure}[t]
\centering
\includegraphics[width=0.55\columnwidth]{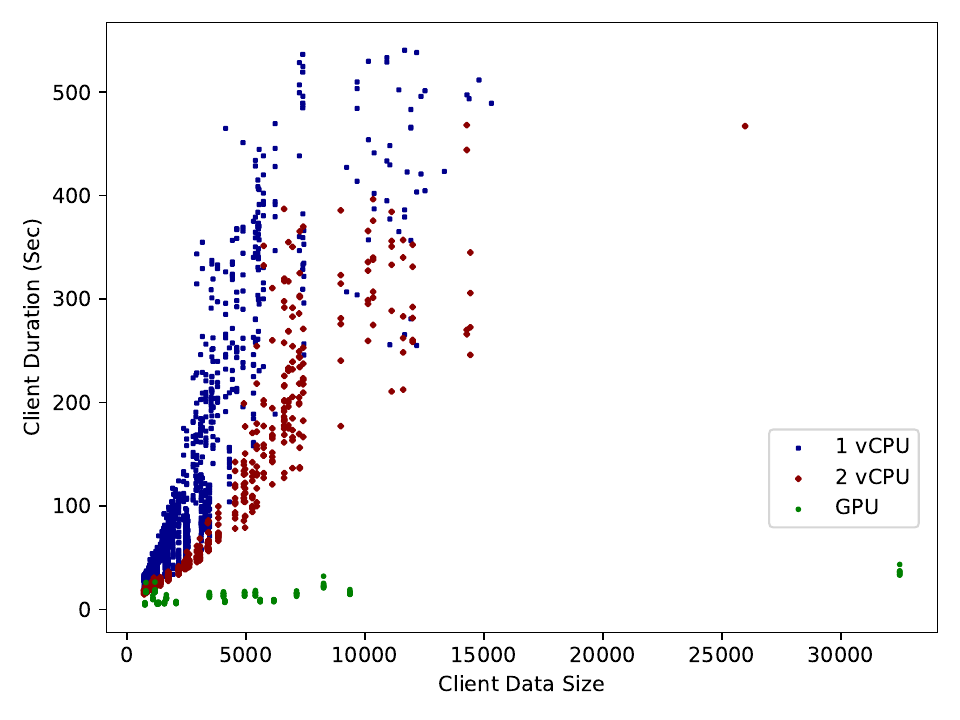}
\vspace{-3mm}
\caption{Client training durations for different hardware resource configurations with non-IID data partitions for the Shakespeare dataset~\cite{caldas2018leaf}.}
\label{fig:clienthardwareduration}
% \shrinkspace
% \vspace{-2mm}
%Better caption
\end{figure}

\begin{algorithm}[t]
  \SetAlgoLined
  \SetKwFunction{calculateScore}{Calculate\_Score}
  \SetKwProg{Pn}{Function}{:}{\Return{$weighted\_average\_score$}}
  \DontPrintSemicolon
  \Pn{\calculateScore{$\beta$, $T$, $N_c$, $E$, $B$}}
  {
  $\#updates \gets \frac{N_c \times E}{B}$

  $weighted\_sum \gets \sum_{i=0}^{k-1} (\lambda^i \times(N_c \times \frac{\#updates}{T_i}))$

  $weighted\_average\_score \gets \beta \times \frac{weighted\_sum}{\sum_{i=0}^{k-1} \lambda^i}$ \tcp*[l]{multiple with $\beta$ to promote clients}
  }

  \caption{Weighted Average Client Score. $\beta$: client's current booster value; $T$: list of training duration for each round; $N_c$: local client data cardinality; $E$: number of local epochs; $B$: client local batch size; $\lambda$: global defined decay rate; $k$: FL training rounds.}
  % \shrinkspace
  \label{alg:calculateScore}
 
\end{algorithm}
 % \vspace{-3mm}
% The $\beta$ is the clients' current booster value, and $T$ is the list of each training duration of the clients, and $N_c$ is the cardinality of the clients' data; $E$ is the number of epochs of the clients, $B$ is the client local batch size, and $\lambda$ is the global defined decay rate.

\begin{algorithm}[t]
  \SetAlgoLined
  \SetKwFunction{selectClients}{Select\_Clients}
  \SetKwProg{Pn}{Function}{:}{\Return{$client\_selection$}}
  \DontPrintSemicolon
  \Pn{\selectClients{$clients$, $clientsPerRound$}}
  {

  Characterize clients as $uninvoked\_clients$ and $invoked\_clients$\;
  Exclude busy clients from \emph{invoked\_clients}\;
  \If{ \#$uninvoked\_clients$ $\geq$ $clientsPerRound$}{
    \Return{Randomly sample $clientsPerRound$ from $uninvoked\_clients$.}
  }
  $client\_scores$ = []\;
  \For{each client in $invoked\_clients$}
  {
    Calculate Weighted Score for $client$. \;
    Append Client Score to $client\_scores$.
  }
  Calculate probability for all \emph{$invoked\_clients$} $\frac{score}{\sum_{}client\_scores}$ \label{alg:line:calc_prob}\;
  $client\_selection$ $\gets$ Sample \emph{$invoked\_clients$} based on probability\;
  Reset booster value for all \emph{clients} in $client\_selection$ \;
  Increase booster value for all \emph{clients} \textbf{NOT} in $client\_selection$ 
  % ($invoked\_clients - client\_selection$)
  }

  \caption{Client selection routine.}
  % \shrinkspace
  \label{alg:client-selection}
\end{algorithm}

\vspace{-3mm}
\subsection{Scoring Clients}
\label{sec:scorebased}
Prior FL training strategies~\cite{chai2020tifl, chai2021fedat},
including \texttt{FedLesScan}~\cite{elzohairy2022fedlesscan} only consider the training duration of clients for selection. This approach is suitable in homogeneous hardware environments, where training time correlates directly with a client's data size. However, this assumption no longer holds true in a heterogeneous setting, as shown in Figure~\ref{fig:clienthardwareduration}. To this end, \texttt{Apodotiko} introduces a scoring-based probabilistic client selection approach considering the client's training duration and data size. 

% our focus shifts to selecting clients that can contribute significantly within a reasonable timeframe.

To score clients, we collect five attributes during each training round: \emph{training duration}, \emph{local client data cardinality}, \emph{batch size}, \emph{number of local epochs}, and a \emph{booster} value. Training duration represents the time required for executing \emph{model.fit} (Algorithm \ref{alg:routine}: Line \ref{alg:line:routine_start_time}-\ref{alg:line:routine_stop_time}), excluding the time spent on network communication and model initialization. Data cardinality represents the total size of a client's local dataset. It is not considered confidential information in FL as it is essential for model aggregation~\cite{mcmahan2017communication} (\S\ref{sec:asyncaggregation}). The booster value is a floating point number used in our strategy for promoting fairness during client scoring and selection.

Algorithm~\ref{alg:calculateScore} describes our strategy for scoring clients. It generates a weighted average score by considering a client's participation in the different FL training rounds. Following this, the calculated scores are then used to select clients that participate in each training round as described in \S\ref{sec:selectingclients}. Evaluating a client's machine learning performance solely based on hardware specifications such as micro-architecture, frequency, and core count can be challenging. Traditional methods often require benchmarking to obtain a score, which can be expensive and inefficient. Towards this, we utilize the client's training performance as a hardware benchmark score, referred to as the \emph{Client Efficiency Score} (CEF). Our scoring algorithm calculates the CEF by determining the number of updates a client makes on the local model. This is done by multiplying the data size ($N_c$) with the epoch size ($E$) and dividing the result by the batch size ($B$) (Line 1). Following this, we calculate the number of updates per second by dividing the total number of updates by the training time (Line 2). A higher CEF score indicates a more powerful hardware resource configuration. Moreover, to account for 
data heterogeneity, we multiply the CEF score with $N_c$. To obtain a score that considers the performance of the client along with the duration of the FL training process, we utilize an exponentially decreasing weighting technique. This technique assigns the highest weight to a client's most recent result ($i=0$) while gradually decreasing the weights for older results (Line 2). The weights are calculated using a globally defined decay rate denoted by $\lambda$. To promote fairness and participation from slow clients in training rounds, we introduce a booster value ($\beta$) that is multiplied by the final score before selection (Line 4). Initially, the booster value is set to 1 for all clients. If a client is available (not busy) (\S\ref{sec:overview}) but not selected for the training round, its booster value is increased by multiplying it with a promotion value greater than one. This ensures that slow clients have a higher probability of being selected in future rounds (\S\ref{sec:selectingclients}). However, if a client is selected, its booster value is reset to one. The decay rate $\lambda$ and promotion rate $\beta$ are determined by an adjustment rate $\rho$ in the range $0 < \rho \le 1$. The adjustment rate controls the extent to which the score weight is increased or decreased. Specifically, the decay rate is calculated as $\lambda = 1 - \rho$, while the promotion rate is calculated as $\beta = 1 + \rho$. By default, the value of $\rho$ is set to 0.2, which is also used in all of our experiments.

\vspace{-2mm}

\subsection{Selecting Clients}
\label{sec:selectingclients}
Algorithm~\ref{alg:client-selection} describes our strategy for client selection. The goal is to sample a given number of clients ($clientsPerRound$) from a list of available clients ($clients$).
Initially, we differentiate between clients who have participated in at least one training round and those who have not yet been invoked (Line 2). Following this, we remove the currently running clients (\S\ref{sec:overview}) from the list of clients that have been invoked once (Line 3). Initially, our algorithm prioritizes uninvoked clients to gather data and enable scoring (\S\ref{sec:scorebased}). If sufficient uninvoked clients are available, the required number of clients is randomly selected from this pool (Lines 4-6).  When the number of required clients exceeds the available uninvoked clients, we 
sample the remaining clients using our scoring strategy (\S\ref{sec:scorebased}). We calculate the score for each available client and append it to the score list. After calculating all scores, we normalize them into values between zero and one. Following this, we transform these normalized scores into probabilities by summing up all scores and dividing each client's normalized score by the total score (Line 12). The higher the client score, the higher the probability, and the more likely the client will be selected for the next round. If a GPU-based client has scored higher than a CPU client, its normalized probability will also be higher, giving it a greater chance of being selected. After obtaining the probability of each client, we randomly sample the required number of clients from the list based on the probability (Line 13). We reset the booster value to 1 for the selected clients so that we do not keep promoting them (Line 14). Finally, we increase the booster value ($\beta$) for the available (\S\ref{sec:overview}) but not selected clients by multiplying it with the promotion rate (Line 15).

% Once we have calculated all scores, we normalize them into values between 0 and 1 to ensure their comparability. In Line \ref{alg:line:calc_prob}, we transform these normalized scores into probabilities by summing up all scores and dividing each client's normalized score by the total score."

% resort to sampling based on the scoring system to select the remaining clients.

% If there are sufficient uninvoked clients available, then the number of required clients is randomly selected from them .

% Our algorithm prioritizes uninvoked clients to gather data and enable scoring.

\vspace{-3mm}

\section{Experimental Results}
\label{sec:expresults}
In this section, we present the performance results for our strategy \texttt{Apodotiko} against other FL training approaches across multiple datasets. For all our experiments, we follow best practices while reporting results.

\vspace{-2mm}

\subsection{Experiment Setup}
\label{sec:expsetup}

\subsubsection{Datasets}
\label{sec:datasets}
For our experiments, we utilize four datasets from various application domains, including image classification, speech recognition, and language modeling, to provide a conclusive evaluation of our strategy's effectiveness. The first dataset we use is the popular handwritten image database called \emph{MNIST}. It contains 60,000 training images and 10,000 images for central evaluation. To simulate a non-IID setting with MNIST, we sort the images by label, split them into 300 shards of 200 images each, and distribute the shards across clients. From the \texttt{LEAF} FL benchmarking framework~\cite{caldas2018leaf}, we utilize the \emph{FEMNIST} and the \emph{Shakespeare} datasets. The 
\emph{FEMNIST} dataset is an extended version of the \emph{MNIST} dataset and contains over 800,000 images. On the other hand, the \emph{Shakespeare} dataset consists of sentences from \emph{The Complete Works of William Shakespeare}, each of length $80$ characters. We employ existing non-IID data partitions available within \texttt{LEAF} for these datasets. From the \texttt{FedScale}~\cite{fedscale} FL benchmark, we utilize the real-world \emph{Google Speech} dataset. This dataset is designed to create simple and useful voice interfaces for applications that use common words like "Yes," "No," and directions. It consists of 105,000 1-second audio files distributed across $2,618$ clients. 

% cThis selection yields an average of approximately 226 samples per client partition for the \emph{FEMNIST} dataset and about 3743 samples per client partition for the \emph{Shakespeare} dataset. 

% Similar to~\cite{elzohairy2022fedlesscan}, we scale down the number of clients to $542$ with each client containing training data of four corresponding clients from \texttt{FedScale}. 

\subsubsection{Model Architectures and Parameters}
\label{sec:modelparams}
For our experiments with the four datasets, we utilize different model architectures that have been used by several previous works in this domain~\cite{caldas2018leaf, serverlessfl, fedless, elzohairy2022fedlesscan, fedscale}. For the \emph{MNIST} dataset, we employ a two-layer Convolutional Neural Network (CNN) with a $5x5$ convolutional kernel. Each convolutional layer is followed by a $2x2$ max pooling layer. The model ends with a fully connected layer with 512 neurons and a ten-neuron output layer. The model comprises $582,026$ trainable parameters in total. Similar to \emph{MNIST}, we use a 2-layer CNN for \emph{FEMNIST}. However, in this case, the network concludes with a fully connected layer comprising $2048$ neurons and an output layer containing $62$ neurons, resulting in $6,603,710$ trainable model parameters. For the \emph{Shakespeare} dataset, we use a Long Short Term Memory (LSTM) recurrent neural network. The model contains an embedding layer of size eight, followed by two LSTM layers with $256$ units and an output layer with a size of $82$. This model has $818,402$ trainable parameters.  The model architecture for the \emph{Google Speech} dataset consists of two identical blocks, followed by an average pooling layer and an output layer with $35$ neurons. Each block consists of two convolutional layers with a $3x3$ convolutional kernel and a max-pooling layer. To prevent overfitting, a dropout layer follows the max-pooling layer, with a rate of $0.25$. In this case, the model has $67,267$ trainable parameters. Across all layers, we use RelU as the activation function, except in the output layer, where the softmax function is utilized. The clients for the \emph{MNIST}, \emph{FEMNIST}, and the \emph{Google Speech} datasets train for five local epochs with a batch size of ten, ten, and five respectively. On the other hand, for the \emph{Shakespeare} dataset, the clients train for one local epoch with a batch size of $32$~\cite{elzohairy2022fedlesscan, fedless}. For the \emph{MNIST}, \emph{FEMNIST}, and the \emph{Google Speech} datasets, we use \texttt{Adam} as the optimizer with a learning rate of $1e-3$. On the other hand, for \emph{Shakespeare}, we use \texttt{SGD} with a learning rate of $0.8$. 

% For the \emph{MNIST} dataset, we use a 2-layer Convolutional Neural Network (CNN) with a $5x5$ kernel. Each convolutional layer is followed by a 2x2 max pooling layer. The model ends with a fully connected layer with 512 neurons and a ten-neuron output layer. In total, the model contains 582,026 trainable parameters. 

\subsubsection{Experiment Configuration}
\label{sec:expconfig}
To effectively scale our experiments and eliminate potential bottlenecks, we set up \emph{FedLess} on a dedicated virtual machine (VM) hosted on our institute's compute cloud. The VM is configured with $40$vCPUs and $177$GiB of RAM. This machine also hosts the file server, providing $200$GiB of storage to accommodate the four datasets utilized in our study. We deployed the aggregator function (\S\ref{sec:serverlessfl}) on a self-hosted, single-node VM Kubernetes (K8s) cluster with OpenFaaS~\cite{openfaas} as the FaaS platform. We configure the VM with $45$GiB of RAM and $10$vCPUs to ensure sufficient resources for efficient operation.

In all our experiments, we deploy the FaaS-based FL clients using the OpenFaaS platform based on K8s. We used OpenFaaS rather than commercial FaaS offerings since it provides us with maximum flexibility for configuring the different clients. In addition, none of the current commercial FaaS offerings support the execution of FaaS functions with GPUs. To enable GPU-based FL clients with OpenFaaS and K8s, we use the 4paradigm's K8s \texttt{vGPU scheduler}~\cite{4paradigmk8svgpuscheduler}. Standard K8s does not support fine-grained allocation or the sharing of GPUs, often leading to underutilization. In contrast, the \texttt{vGPU scheduler}  balances GPU usage across nodes and allows users to allocate resources based on device memory and core usage, thereby increasing GPU utilization.

Across all datasets (\S\ref{sec:datasets}), we use $200$ clients and sample $100$ clients per round unless otherwise specified. We configure $130$ clients with $1$vCPU
and a memory limit of $2048$MiB, $50$ clients with $2$vCPUs and a memory limit of $4096$MiB, and an additional $20$ clients spread across five Nvidia P100 GPUs, each configured with $0.4$ vGPU~\cite{4paradigmk8svgpuscheduler}. This diverse client setup, ranging from CPU to GPU configurations, allows for a comprehensive exploration of our strategy's behavior and performance in heterogeneous environments, offering insights into its practical efficacy in serverless FL systems.

% The default \emph{Kubernetes} architecture does not support GPU splitting or fine-grained allocation and sharing of GPU resources, which makes assigning GPU resources as floating points challenging. \cite{k8s.gpu}
% To address this issue, plugins like \emph{KubeShare} \cite{10.1145/3369583.3392679} and 4paradigm's k8s vGPU scheduler \cite{4paradigmk8svgpuscheduler} have been developed. \emph{KubeShare} extends \emph{Kubernetes} to allow the allocation and sharing of GPUs as first-class resources for scheduling. However, at the time of writing, \emph{KubeShare} did not work, and we opted for the \emph{4paradigm k8s vGPU scheduler}, which enables GPU sharing among tasks, device memory control, virtual device memory, and GPU type specification.
% The vGPU scheduler balances GPU usage across nodes and allows users to allocate resources based on device memory and core usage, thereby increasing GPU utilization. Although the vGPU scheduler is based on the \emph{NVIDIA device plugin} and retains its official features, it lacked proper limitations to GPU memory on pods at the time of writing. To address this issue, we restricted memory allocation at the \emph{TensorFlow} level, ensuring that one client would not occupy all memory on the GPU.

\subsubsection{Baseline Strategies}
\label{sec:baselinestrategies}
To effectively evaluate \texttt{Apodotiko}, we compare it against five other strategies: \texttt{FedAvg}~\cite{mcmahan2017communication}, \texttt{FedProx}~\cite{li2020federated}, \texttt{SCAFFOLD}~\cite{scaffold}, \texttt{FedLesScan}~\cite{elzohairy2022fedlesscan}, and \texttt{FedBuff}~\cite{fedbuff}. \texttt{FedProx} and \texttt{SCAFFOLD} are significantly popular conventional FL strategies in heterogeneous environments, while \texttt{FedBuff} is utilized in production at Meta~\cite{huba2022papaya}. For \texttt{Apodotiko}, we fix the \emph{concurrencyRatio} to $0.3$ in all our experiments unless otherwise specified.

\begin{figure*}[t]
\begin{subfigure}{\textwidth}
    \centering
        \includegraphics[width=0.24\columnwidth]{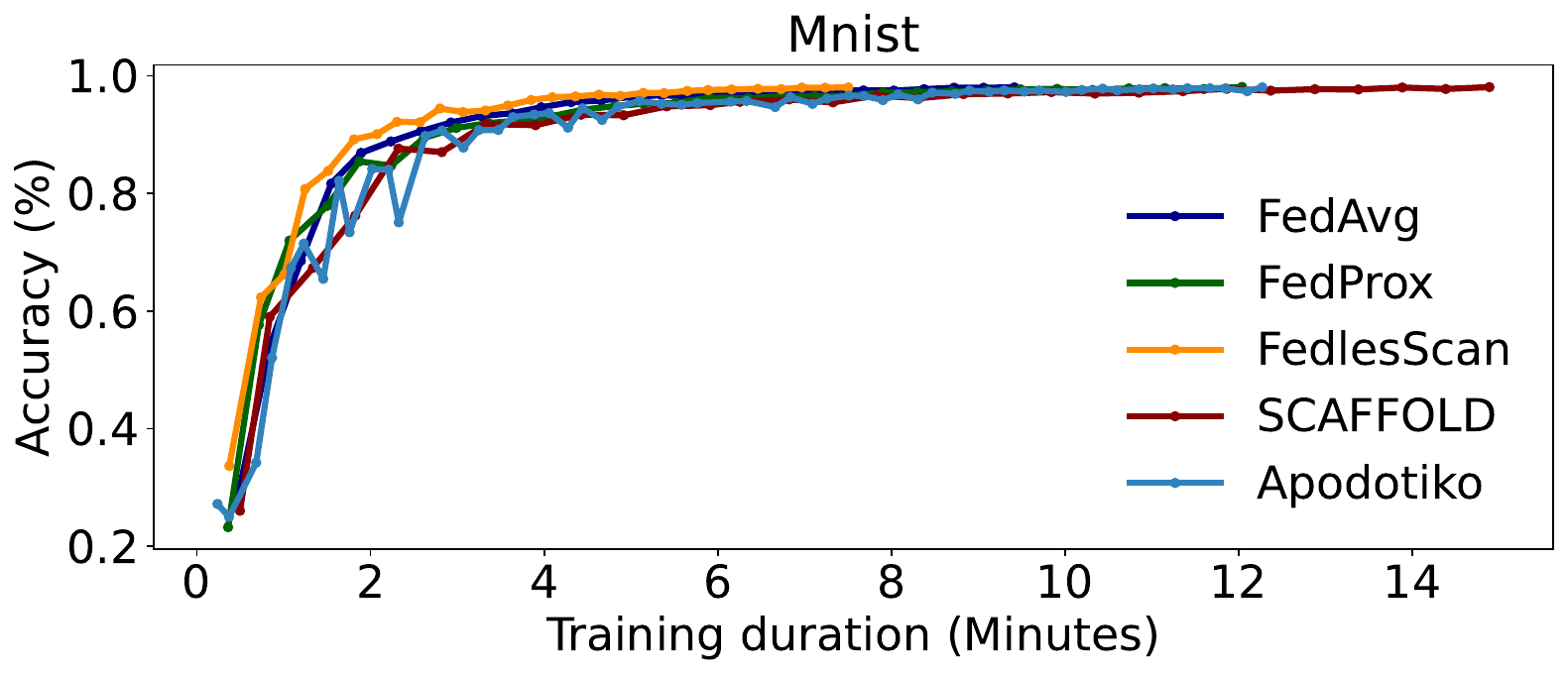}
        \includegraphics[width=0.24\columnwidth]{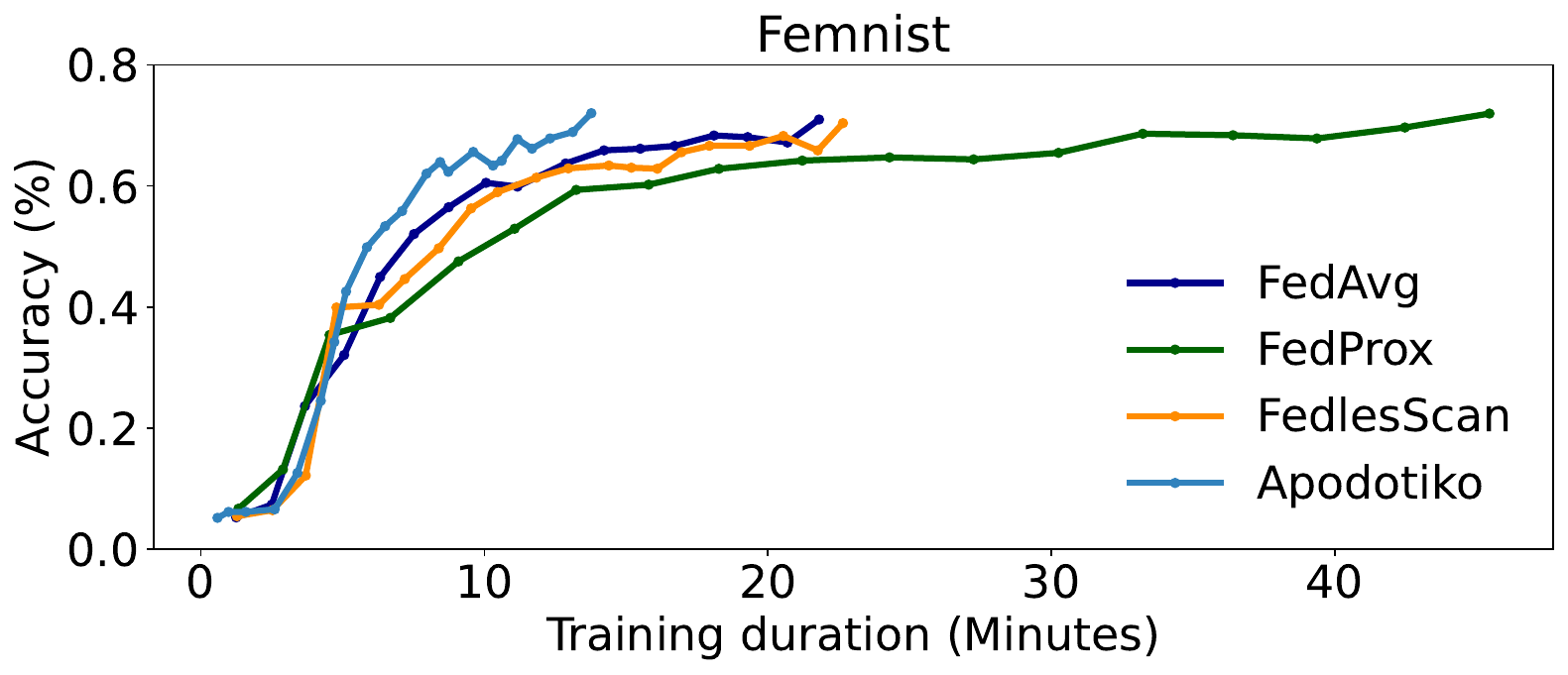}
        \includegraphics[width=0.24\columnwidth]{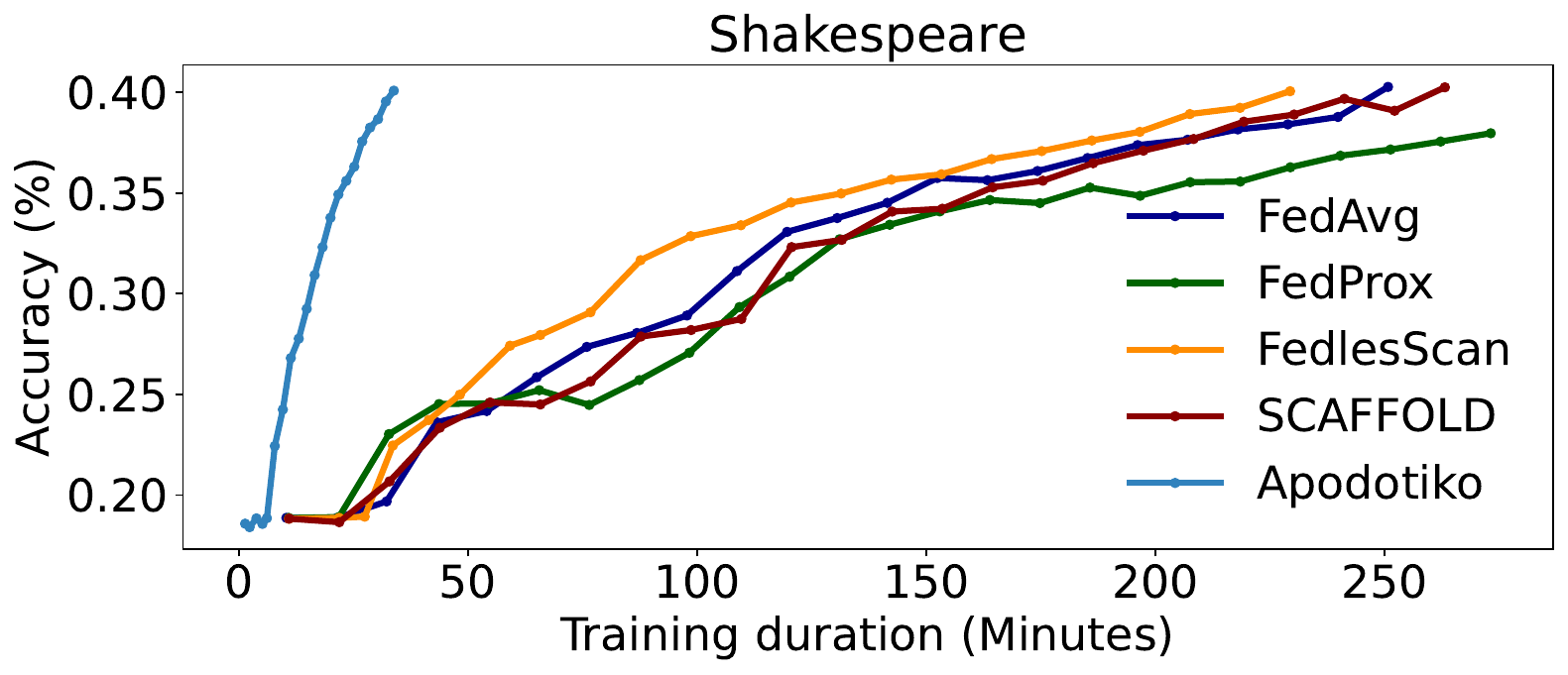}
        \includegraphics[width=0.24\columnwidth]{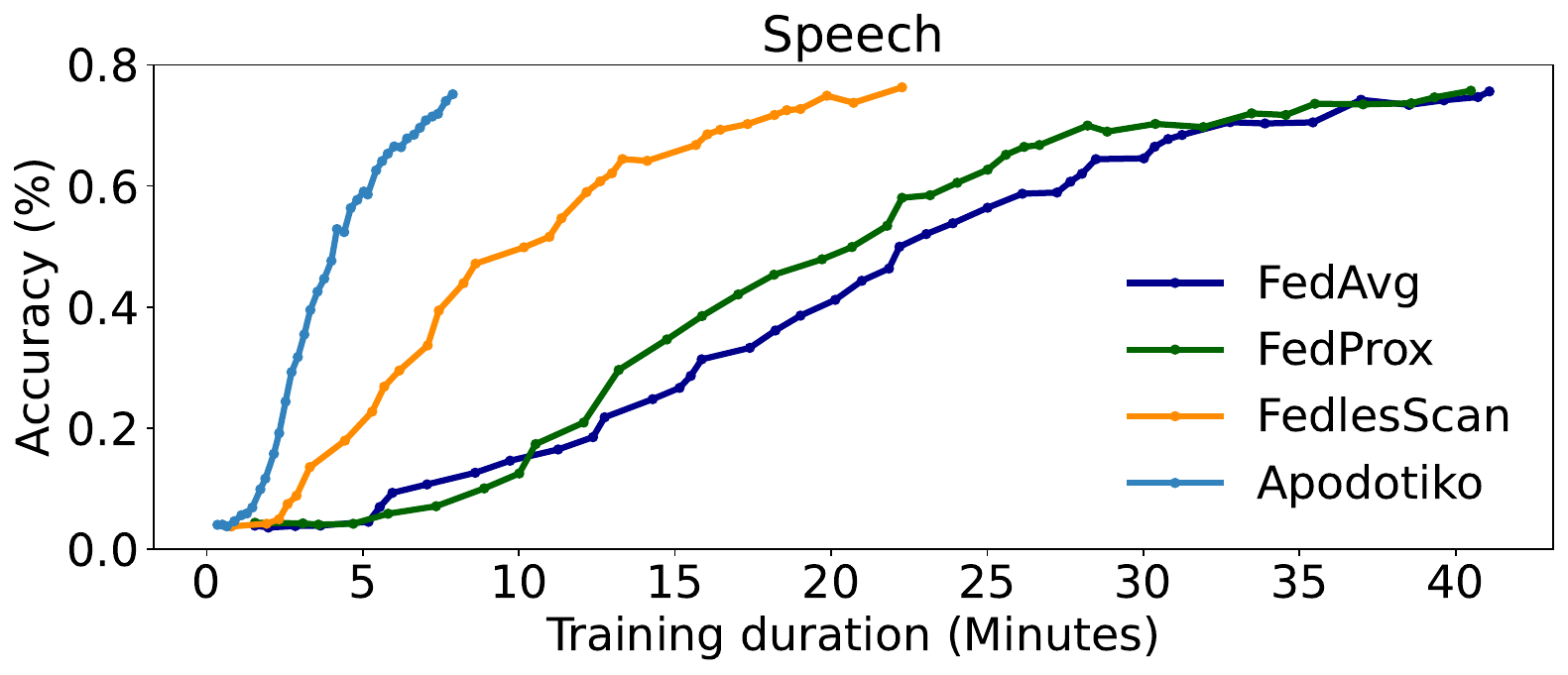}
        \caption{Comparing \emph{model accuracy}.}
        \label{fig:comparingaccfinal}
\end{subfigure}
\begin{subfigure}{\textwidth}
    \centering
   \includegraphics[width=0.24\columnwidth]{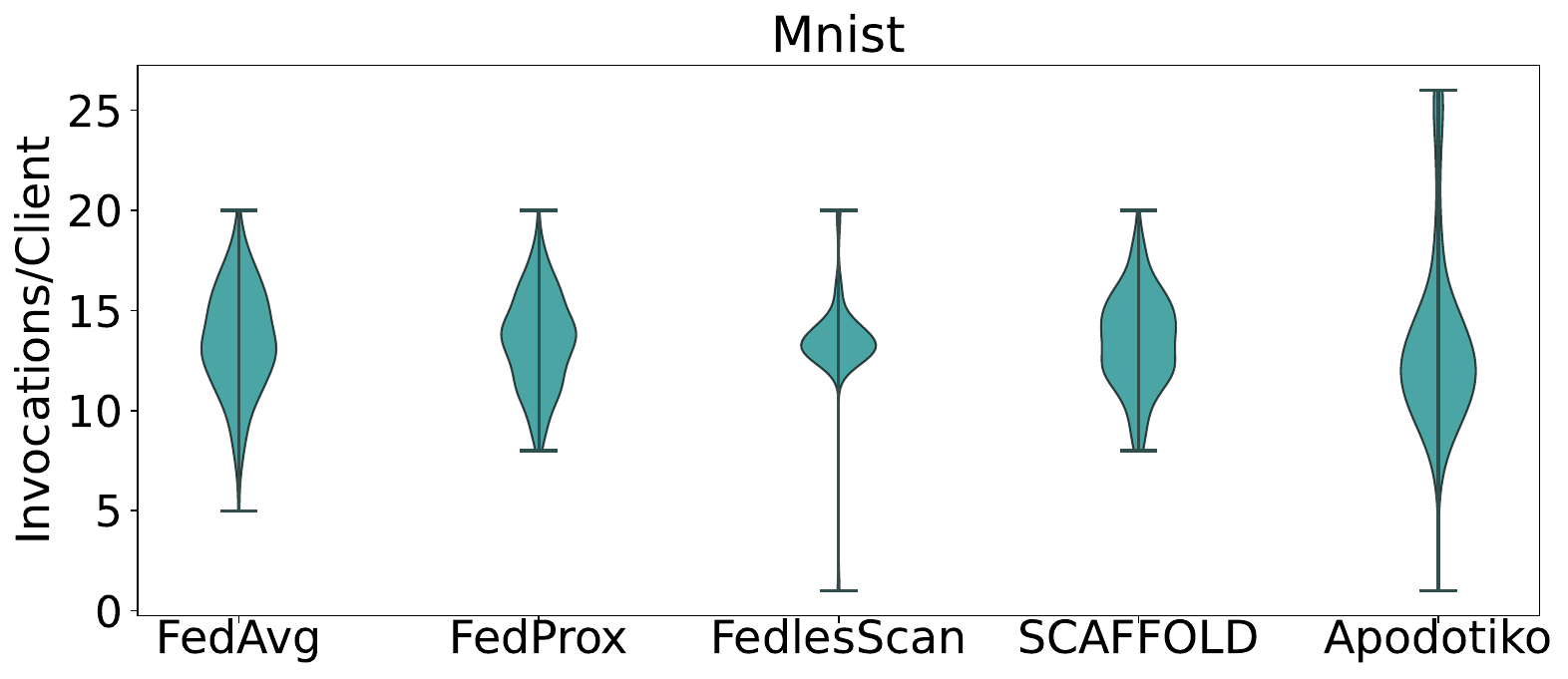}
   \includegraphics[width=0.24\columnwidth]{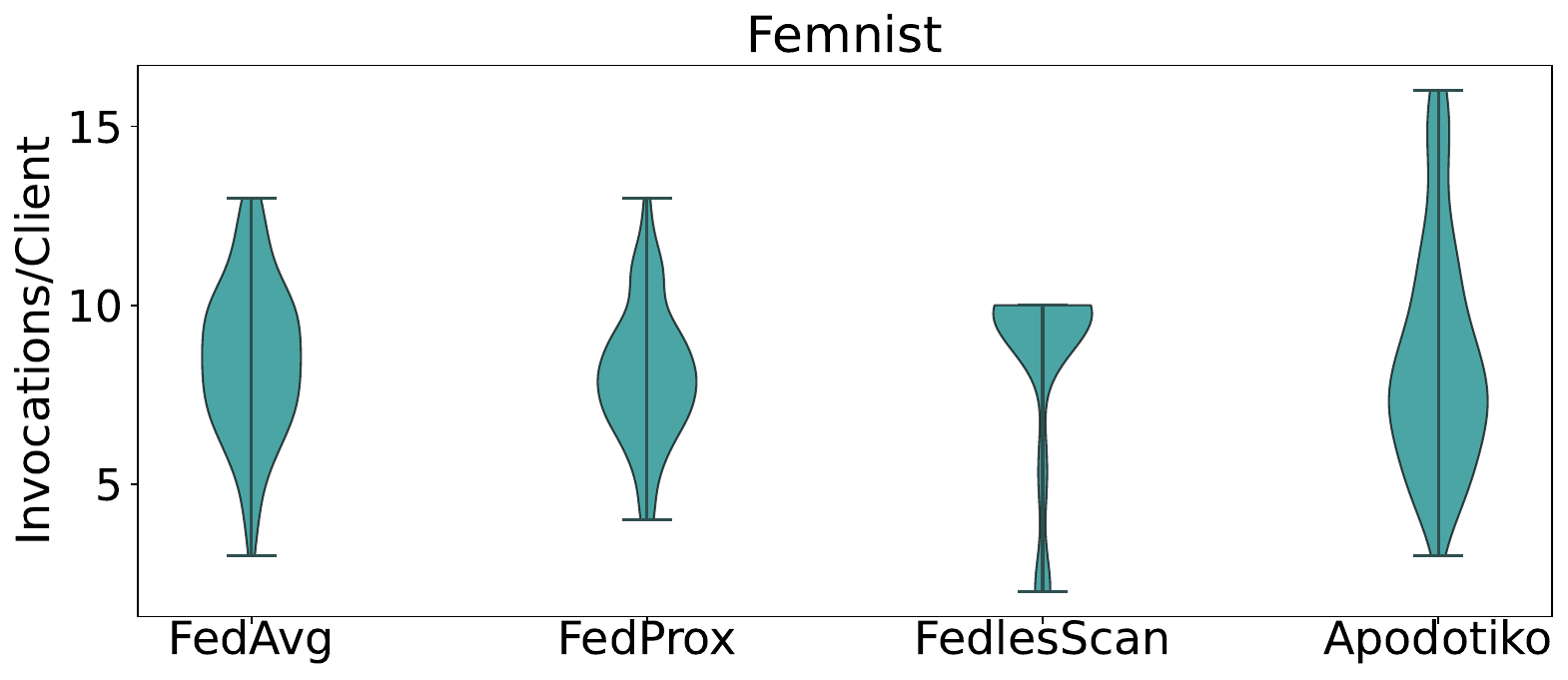}
   \includegraphics[width=0.24\columnwidth]{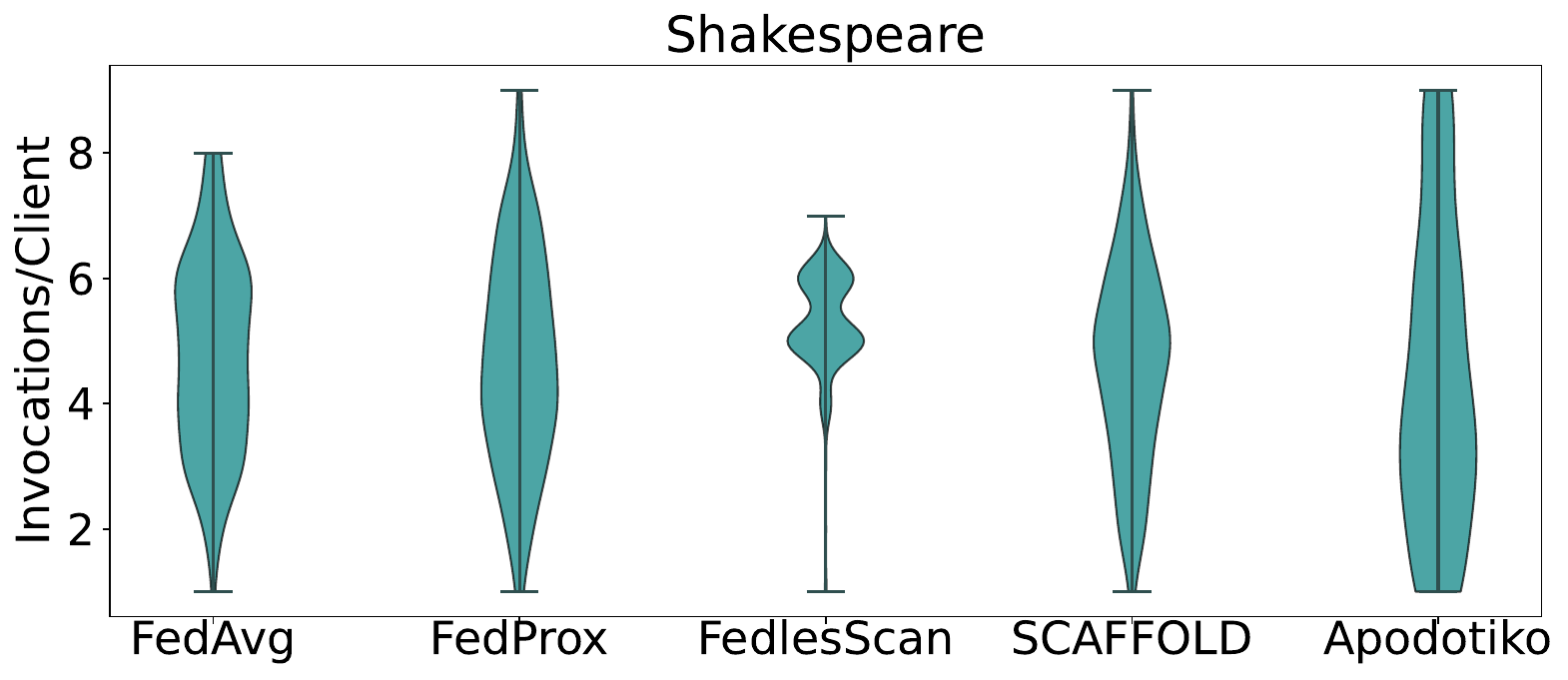}
    \includegraphics[width=0.24\columnwidth]{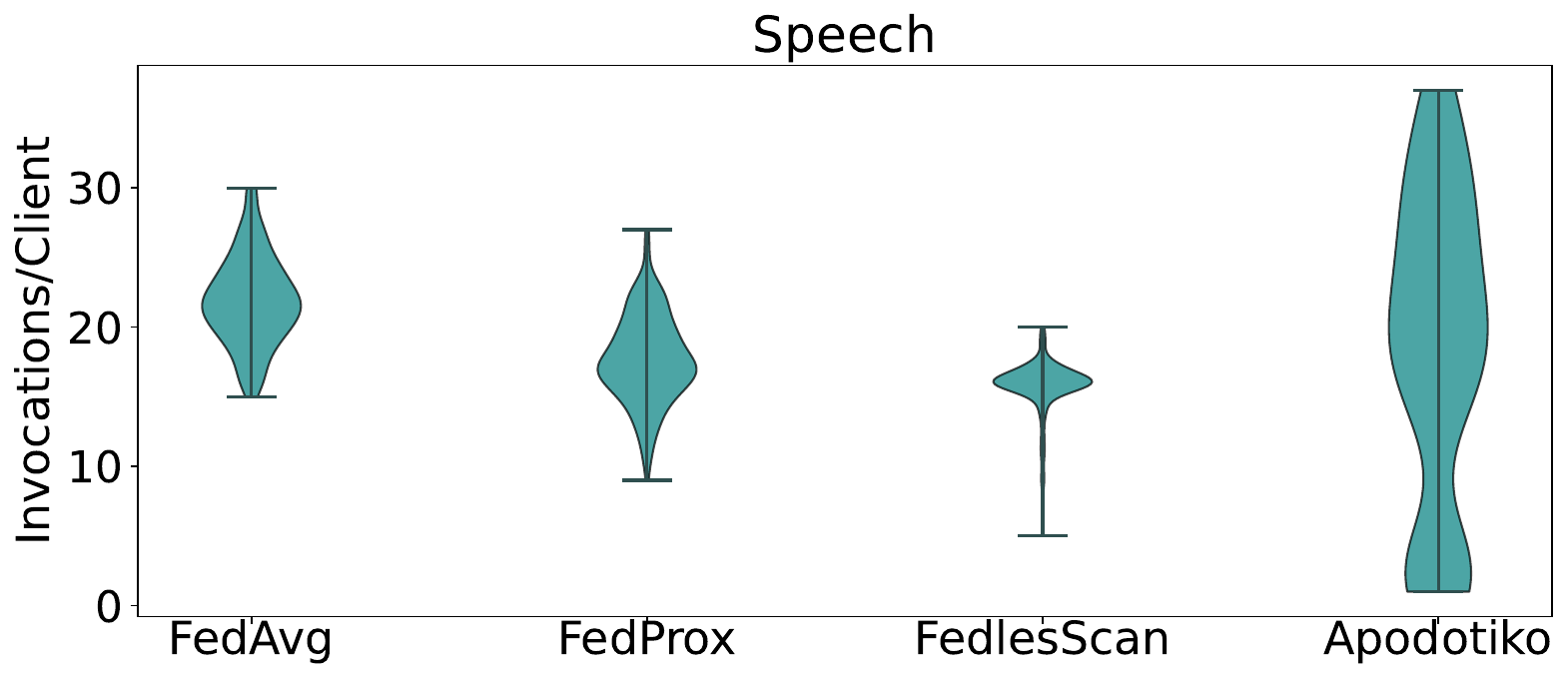}
        \caption{Comparing \emph{client selection bias}.}
        \label{fig:compbiasfinal}
\end{subfigure}
\begin{subfigure}{\textwidth}
    \centering
         \includegraphics[width=0.24\columnwidth]{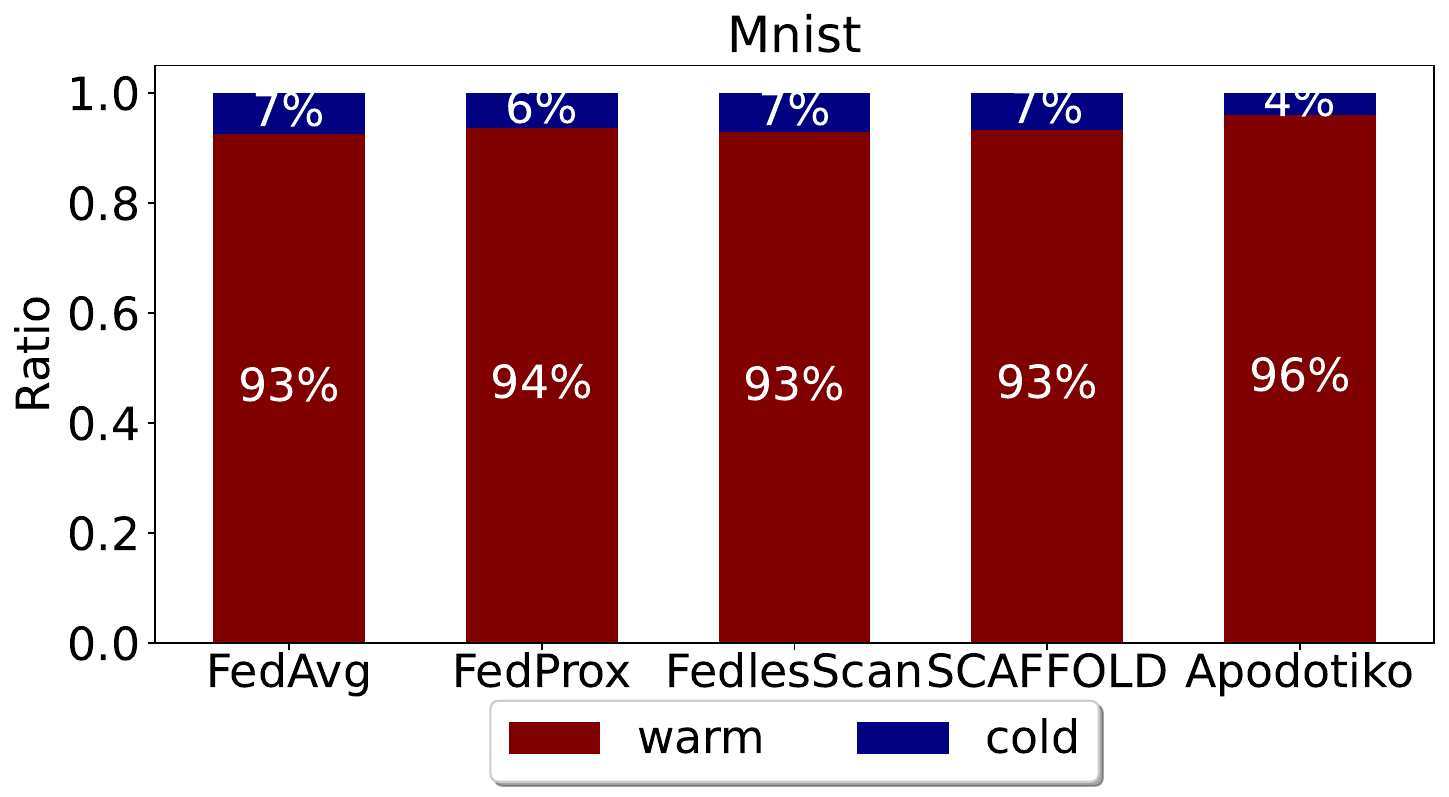}
   \includegraphics[width=0.24\columnwidth]{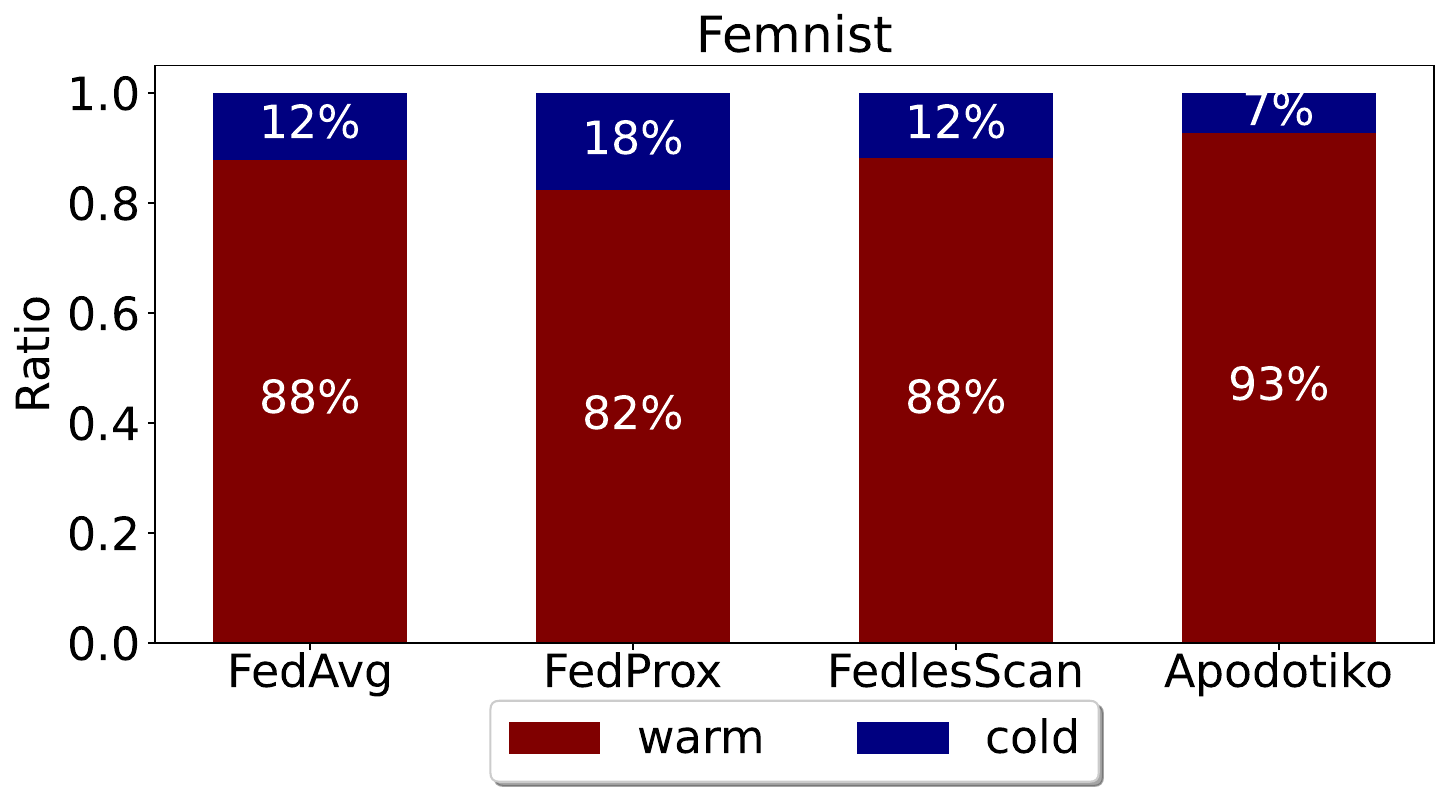}
   \includegraphics[width=0.24\columnwidth]{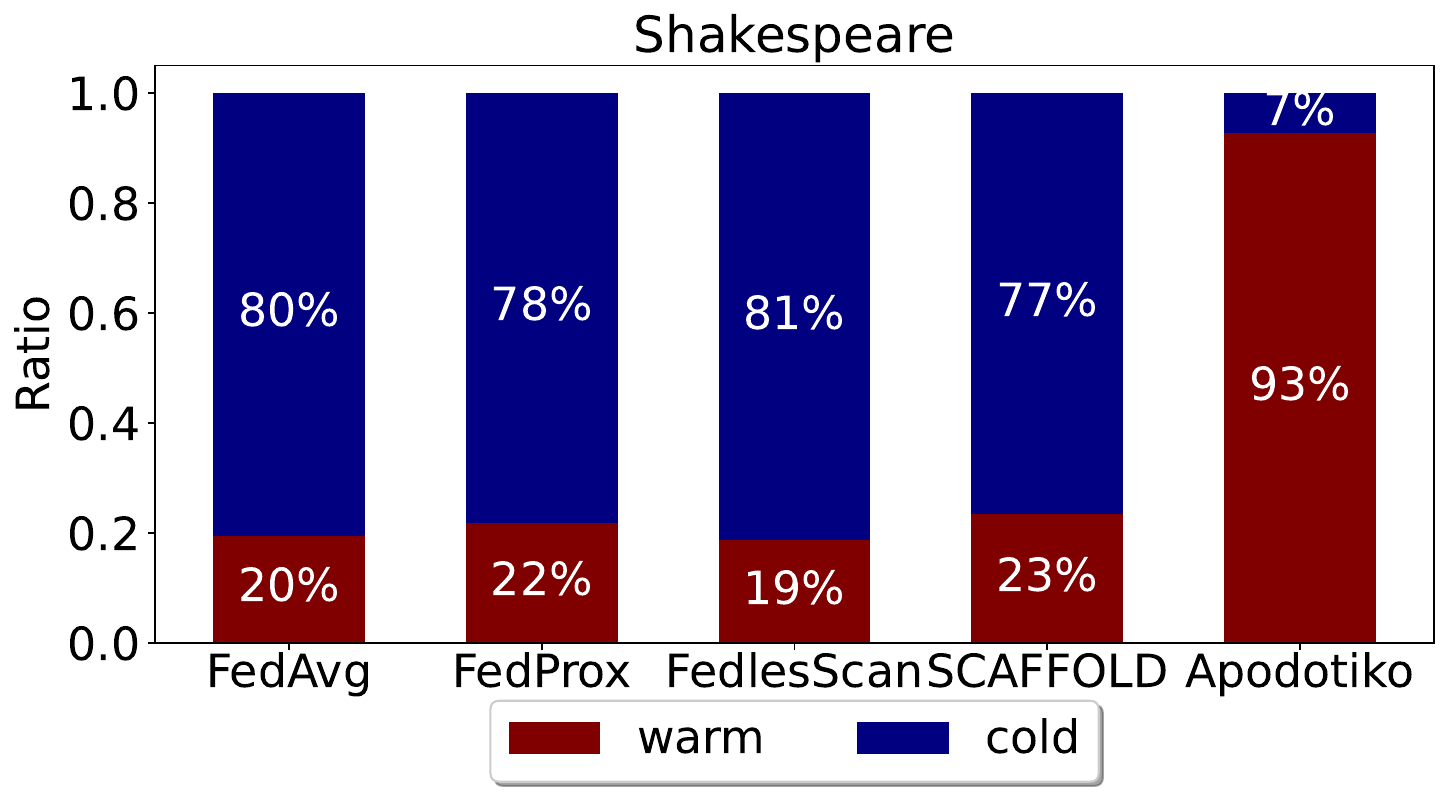}
    \includegraphics[width=0.24\columnwidth]{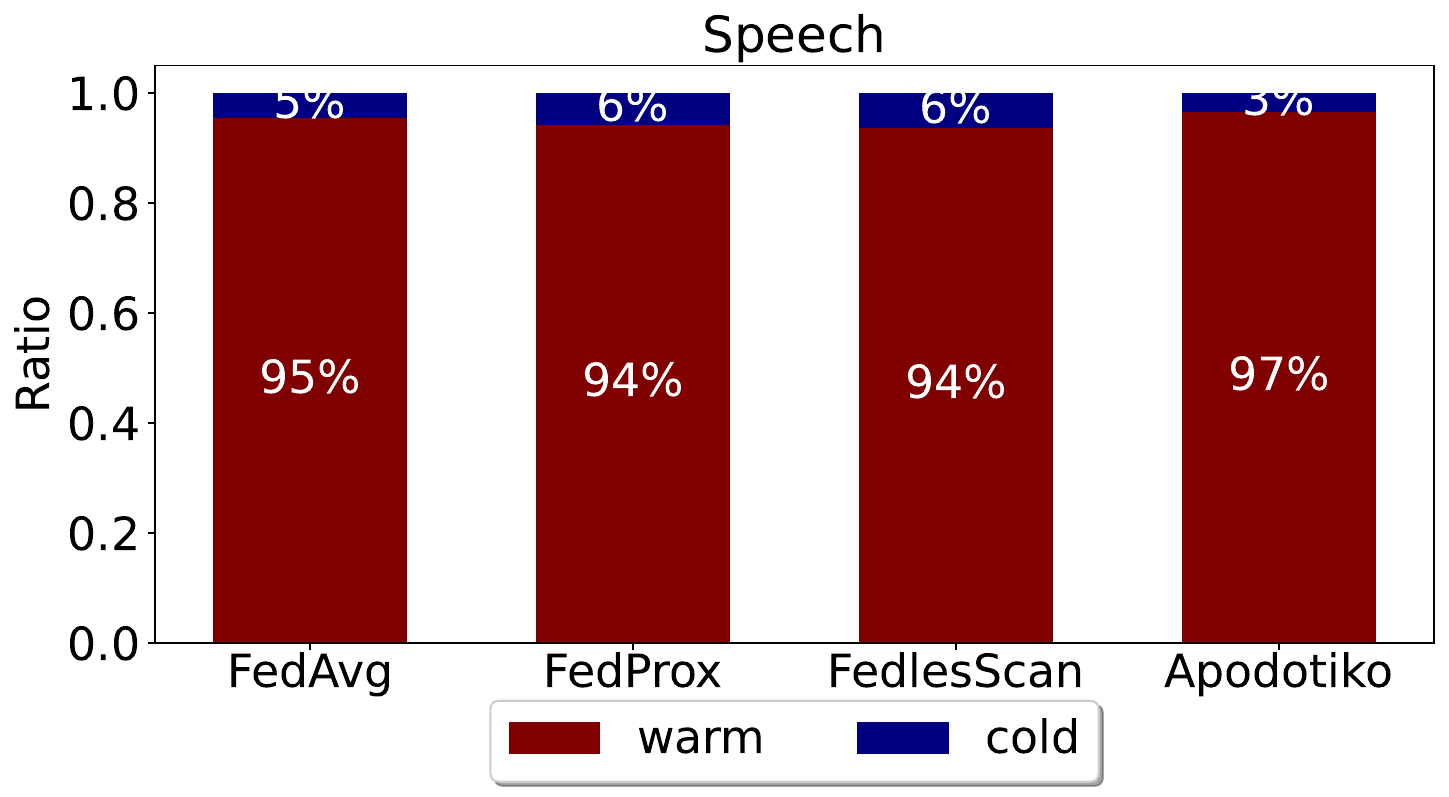}
        \caption{Comparing \emph{cold start ratios}.}
        \label{fig:compcoldstartfinal}
\end{subfigure}
% \hspace{-5mm}
\vspace{-5mm}
\caption{Comparing different evaluation metrics across the different FL strategies.}
\label{fig:accmetricscomp}
\shrinkspace
%Better caption
\end{figure*}

\begin{figure}
\begin{subfigure}{0.23\textwidth}
    \centering
        \includegraphics[width=\columnwidth]{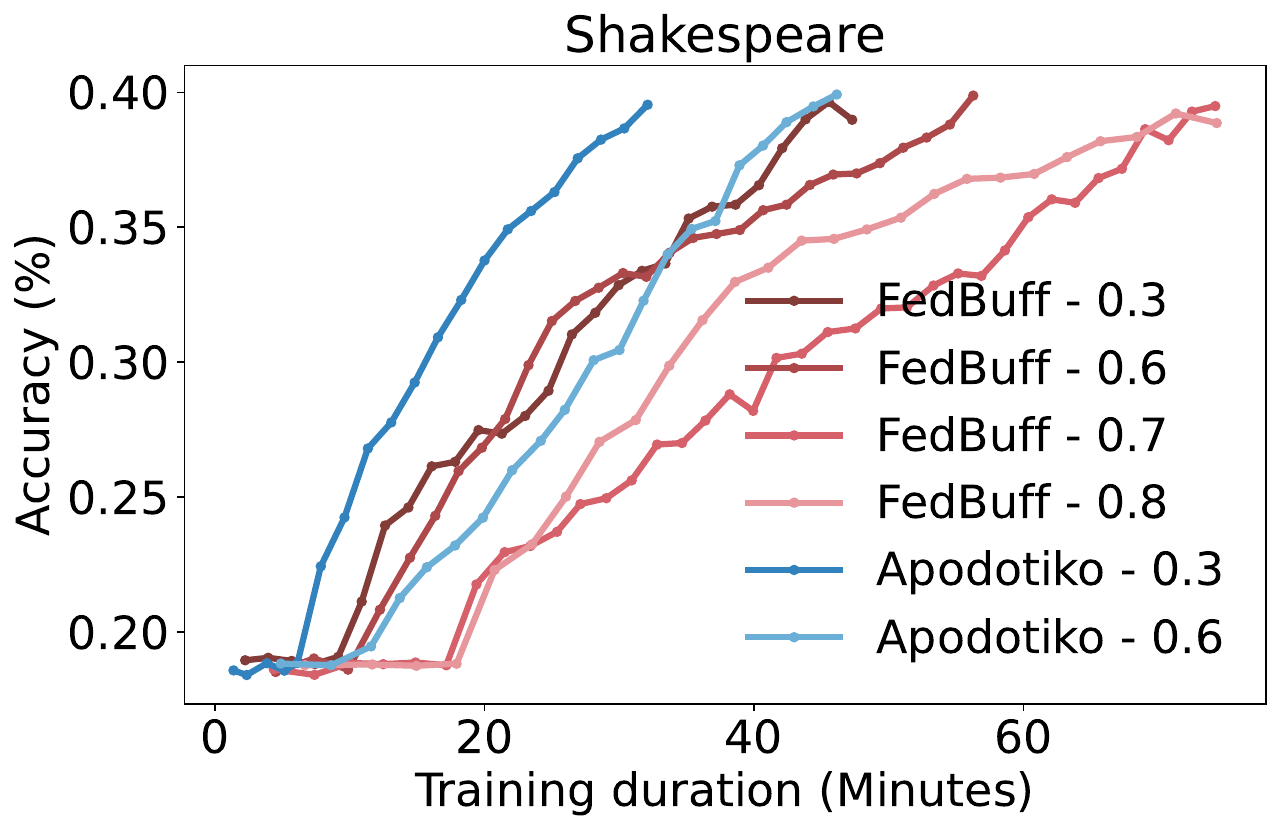}
        \caption{\emph{Shakespeare}.}
        \label{fig:fedbuffshakes}
\end{subfigure}
\begin{subfigure}{0.23\textwidth}
    \centering
        \includegraphics[width=\columnwidth]{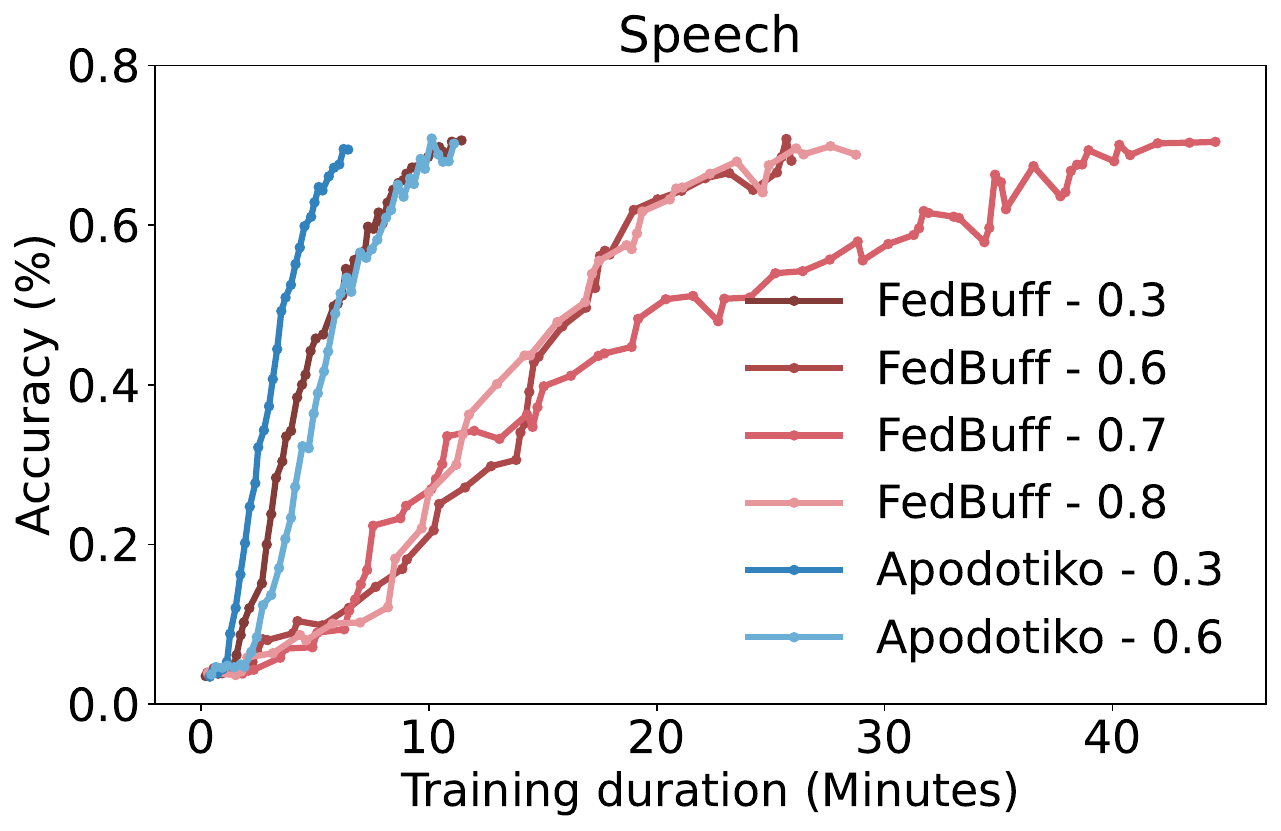}
        \caption{\emph{Google Speech}.}
        \label{fig:fedbuffspeech}
\end{subfigure}
% \hspace{-5mm}
\vspace{-2mm}
\caption{Comparing \texttt{Apodotiko} with \texttt{FedBuff}.}
\label{fig:fedbuffcomp}
%Better caption
% \shrinkspace
\end{figure}

\subsubsection{Evaluation Metrics}
\label{sec:evalmetrics}
For comparing the different strategies (\S\ref{sec:baselinestrategies}), we use metrics that cover four aspects: \emph{model performance}, \emph{client selection bias}, \emph{cold start ratio}, and \emph{strategy efficiency}. To evaluate model performance, we calculate model accuracy and track the accuracy progress throughout the FL training process. To ensure a fair evaluation of the newest global model, we use a distributed evaluation approach. Towards this, we randomly select clients after each FL training round to evaluate the global model on their test datasets. Following this, we calculate a weighted average of the obtained accuracy values from the different clients to obtain the final model accuracy. Client selection bias represents the variations in client invocations during the FL training process, offering insights into the effectiveness of the selection strategy. We quantify bias by calculating the difference between the least-called and most-called clients~\cite{wu2020safa, elzohairy2022fedlesscan}. Low bias is desirable for scenarios with minimal stragglers, while prioritizing reliable clients may be necessary for straggler-heavy environments, leading to increased bias. A key characteristic of the FaaS computing model is \texttt{scale-to-zero}, i.e., idle function instances are automatically terminated if there are no function invocation requests within a given time frame. In serverless FL, this can cause cold starts for client function instances, leading to increased training durations and expenses~\cite{fedless, elzohairy2022fedlesscan, jayaram2022lambda}. To compute the cold start ratio, we calculate the total number of client cold start invocations and divide it by the total number of invocations across all clients. We differentiate between cold and warm start invocations by monitoring the client function instances in our K8s cluster. In our experiments, we configured the function instances to scale down after remaining inactive for ten minutes. To assess the efficiency of our strategy, we calculate the total training time and costs. The total time represents the duration required to achieve the target global model accuracy. For estimating costs for our experiments, we use the cost model provided by GCP~\cite{gcp.VM.pricing, gcp.function.pricing}. or CPU-based clients, we calculate costs based on the allocated memory, CPU, and function duration. In contrast, for GPU clients, we calculated costs by considering the hourly rate of the GPU model and the proportion of GPU resources utilized during the training process.

\vspace{-3mm}

\subsection{Comparing Accuracy}
\label{sec:compacc}
In this subsection, we focus on demonstrating the improved convergence rate of \texttt{Apodotiko} as compared to other FL strategies rather than pursuing state-of-the-art model accuracies on these tasks. Towards this, we limit target model accuracies to $0.98$ for \emph{MNIST}, $0.70$ for \emph{FEMNIST}, $0.40$ for \emph{Shakespeare}, and $0.75$ for the \emph{Google Speech} dataset. Figure~\ref{fig:comparingaccfinal} illustrates the model accuracies throughout the FL training process across various strategies. For the \emph{MNIST} dataset, we observe that all strategies have similar performance, while \texttt{FedlesScan} sightly outperforms other training strategies. This can be attributed to two reasons. First, \emph{MNIST} is a relatively small dataset with low training iteration durations. Second, \emph{MNIST}'s balanced non-IID data distribution allows \texttt{FedLesScan} to benefit from its client clustering strategy based solely on training times. On the other hand, for the \emph{FEMNIST} dataset, we observe that \texttt{FedAvg} performs better than \texttt{FedlesScan} due to unbalanced non-IID data distributions and increased training times. \texttt{Apodotiko} outperforms all other strategies and achieves the target model accuracy with a speedup of 1.73x compared to \texttt{FedAvg}. For the \emph{Shakespeare} dataset, we observe a significant performance difference across the different training strategies. \texttt{Apodotiko} outperforms all other strategies with a speedup of $7$x as compared to \texttt{FedAvg}, $6.63$x against \texttt{FedLesScan}, $7.2$x against \texttt{SCAFFOLD}, and $7.82$x against \texttt{FedProx}. The non-IID data partitions provided by LEAF for the \emph{Shakespeare} dataset introduce significant variations in training durations across different hardware resource configurations, as shown in Figure~\ref{fig:clienthardwareduration}. This variance can lead to situations where clients with limited resources might fail to complete their training before a FL round ends, resulting in wasted contributions. However, our client selection strategy (\S\ref{sec:scorebased}), driven by scoring based on data size and hardware resources, prioritizes clients with larger data sizes and more powerful hardware, ensuring greater contributions to the global model. Moreover,
our asynchronous aggregation technique with a stateless weighting function (\S\ref{sec:asyncaggregation}) accommodates late contributions, leading to better global model accuracy. These two factors combined contribute to the superior performance of \texttt{Apodotiko}. Similarly, for the \emph{Google Speech} dataset, \texttt{Apodotiko} achieves a speedup of $6.19$x compared to \texttt{FedAvg} and $3.3$x compared to \texttt{FedLesScan}. In our experiments, \texttt{SCAFFOLD} did not converge for the \emph{FEMNIST} and the \emph{Speech} datasets. As a result, we omit it in Figure~\ref{fig:accmetricscomp}.

While Figure~\ref{fig:comparingaccfinal} compares \texttt{Apodotiko} with \emph{synchronous} and \emph{semi-asynchronous} strategies, Figure~\ref{fig:fedbuffcomp} presents a comparative analysis with the \emph{asynchronous} strategy \texttt{FedBuff}~\cite{fedbuff} for the \emph{Shakespeare} and \emph{Google Speech} datasets. For \texttt{FedBuff}, we vary the buffer ratio from $0.3$ to $0.8$, while for our strategy, we present results for the \emph{concurrencyRatios} (CR) of $0.3$ and $0.6$. For the \emph{Shakespeare} dataset, our strategy using CRs of $0.3$ and $0.6$ outperforms \texttt{FedBuff} with a buffer ratio of $0.3$, achieving a speedup of $1.43$x and $1.06$x respectively. Similarly, for the \emph{Google Speech} dataset, our
strategy achieves a speedup of $1.74$x and $1.01$x with CRs of $0.3$ and $0.6$ respectively, compared to \texttt{FedBuff} with a buffer ratio of $0.3$. The better performance of our strategy can be attributed to our intelligent client selection methodology that prioritizes clients that contribute more to the global model, in contrast to \texttt{FedBuff}, which selects clients randomly.

.

\vspace{-6mm}

\subsection{Comparing Client Selection Bias and Cold Start Ratios}
\label{sec:compbiasacsr}
Figure~\ref{fig:compbiasfinal} presents insights into the client selection biases among various FL training strategies across multiple datasets. Using violin plots, we visualize a distribution based on the number of invocations for each client (y-axis). The height difference between the highest and lowest points in the distribution represents the degree of bias. A greater height indicates a stronger bias towards a specific subset of clients, while a lower height suggests a more balanced distribution of invocations. Moreover, a wider width in the distribution indicates that certain clients were invoked more frequently.  Across all datasets, we observe that \texttt{FedAvg}, \texttt{FedProx}, and \texttt{SCAFFOLD} show normal distributions due to the random selection of clients. For the \emph{MNIST} dataset, we observe that \texttt{FedLesScan} appears more centralized, indicating a balanced allocation of training among clients, with only a few outliers. In contrast, our strategy shows a relatively normal distribution of client invocations, with most clients being invoked between five and $15$ times. However, the line stretches out more than other strategies, with some clients receiving over $25$ invocations and few with as little as five. As the \emph{MNIST} dataset involves balanced non-IID data distributions among clients, the scoring strategy in \texttt{Apodotiko} primarily differentiates based on the training time, which is mainly influenced by clients' hardware resources. As a result, GPU-based clients receive more frequent invocations than other clients. For the \emph{FEMNIST} dataset, we observe that \texttt{FedLesScan} is more concentrated in the middle as it prioritizes clients with lower training durations.  Moreover, it overlooks clients with larger data sizes and limited computational resources, often characterizing them as stragglers and selecting them less frequently in training. In contrast, our strategy maintains a relatively balanced distribution due to the probabilistic client selection approach described in \S\ref{sec:selectingclients}. This method ensures that every client retains a chance of being selected, even those with longer training times. For the \emph{Shakespeare} dataset, \texttt{FedLesScan} shows a distribution similar to that for the \emph{FEMNIST} dataset. On the other hand, \texttt{Apodotiko} exhibits a slightly fatter end on the bottom compared to the \emph{FEMNIST} dataset. This distribution arises from the significant differences in training durations observed across various hardware resource configurations, as shown in Figure~\ref{fig:clienthardwareduration} (\S\ref{sec:compacc}). For the \emph{Google Speech} dataset, \texttt{FedLesScan} shows a normal distribution, similar to the results for the \emph{MNIST} dataset, but with lower variance compared to both \texttt{FedAvg} and \texttt{FedProx}.
In contrast, our strategy portrays a distribution distinct from the one seen in the \emph{MNIST} dataset. This divergence stems from the unbalanced non-IID data distribution in the \emph{Google Speech} dataset, causing client data size to hold a more significant role in the scoring process.

\begin{table}[t]
  \centering
  \begin{adjustbox}{width=0.9\columnwidth,  center}
    \begin{tabular}{|cc|rc|rc|rc|rc|}
      \hline
      \multicolumn{2}{|c|}{\diagbox{Strategy}{Dataset}}  & \multicolumn{2}{c|}{\textbf{MNIST (min)}} & \multicolumn{2}{c|}{\textbf{FEMNIST (min)}} & \multicolumn{2}{c|}{\textbf{Shakespeare (min)}} & \multicolumn{2}{c|}{\textbf{Speech (min)}}                                                                                                   \\ \hline
      \multicolumn{2}{|c|}{FedAvg~\cite{mcmahan2017communication}}                      & 10.98                               & (1.00x)                               & 22.44                                     & (1.00x)                              & 245.98           & (1.00x)                & 49.78            & (1.00x)                          \\ \hline
      \multicolumn{2}{|c|}{FedProx~\cite{li2020federated}}                      & 15.03                               & (0.73x)                               & 39.46                                     & (0.57x)                              & 273.58           & (0.90x)                & 53.20            & (0.94x)                          \\ \hline
      \multicolumn{2}{|c|}{FedLesScan~\cite{elzohairy2022fedlesscan}}                   & \textbf{9.69}                       & (\textbf{1.13x})                      & 25.88                                     & (0.87x)                              & 232.18           & (1.06x)                & 26.59            & (1.87x)                          \\ \hline
      \multicolumn{2}{|c|}{SCAFFOLD~\cite{scaffold}}                 & 14.31                               & (0.77x)                               & \multicolumn{2}{r|}{-}                    & 252.07                               & (0.98x)          & \multicolumn{2}{r|}{-}                                                       \\ \hline
      \multicolumn{1}{|c|}{\multirow{4}{*}{Apodotiko}} & CR = 0.3                                 & 11.83                                 & (0.93x)                                   & \textbf{12.95}                       & \textbf{(1.73x)} & \textbf{34.98}         & \textbf{(7.03x)} & \textbf{8.04} & \textbf{(6.19x)} \\ \cline{2-10}
      \multicolumn{1}{|c|}{}                             & CR = 0.6                                 & 11.65                                 & (0.94x)                                   & 18.28                                & (1.23x)          & 69.04                  & (3.56x)          & 12.18         & (4.09x)          \\ \cline{2-10}
      \multicolumn{1}{|c|}{}                             & CR = 0.7                                 & 12.21                                 & (0.90x)                                   & 20.21                                & (1.11x)          & 51.28                  & (4.80x)          & 12.74         & (3.91x)          \\ \cline{2-10}
      \multicolumn{1}{|c|}{}                             & CR = 0.8                                 & 10.92                                 & (1.01x)                                   & 22.72                                & (0.99x)          & 72.66                  & (3.39x)          & 16.42         & (3.03x)          \\ \hline
    \end{tabular}%
  \end{adjustbox}
  \caption{Comparing total training duration (min) across various FL strategies and datasets. The highlighted values represent the best-performing strategy for a particular dataset.}
  \label{tab:duration-comparison}
  \shrinkspace
\end{table}

\begin{figure}[t]
\begin{subfigure}{0.23\textwidth}
    \centering
        \includegraphics[width=\columnwidth]{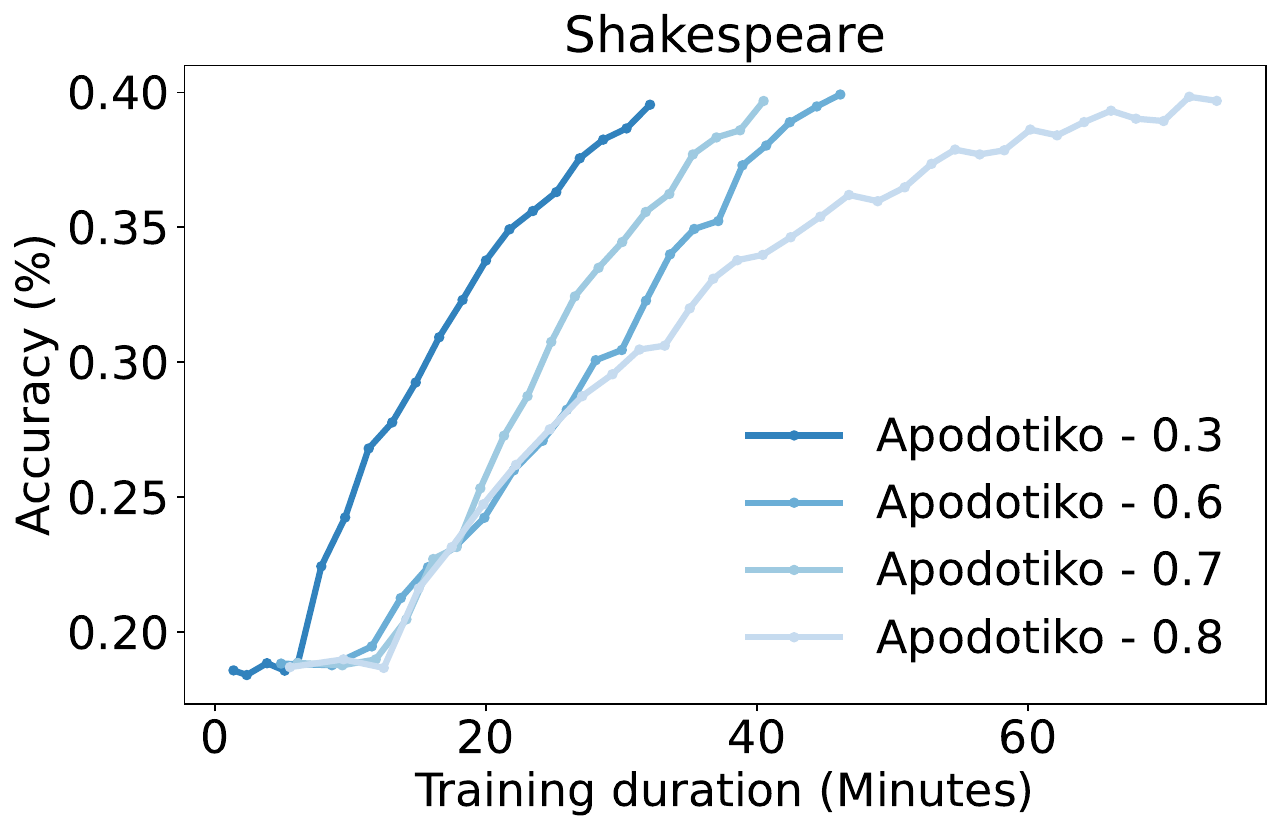}
        \caption{\emph{Shakespeare}.}
        \label{fig:buffshakes}
\end{subfigure}
\begin{subfigure}{0.23\textwidth}
    \centering
        \includegraphics[width=\columnwidth]{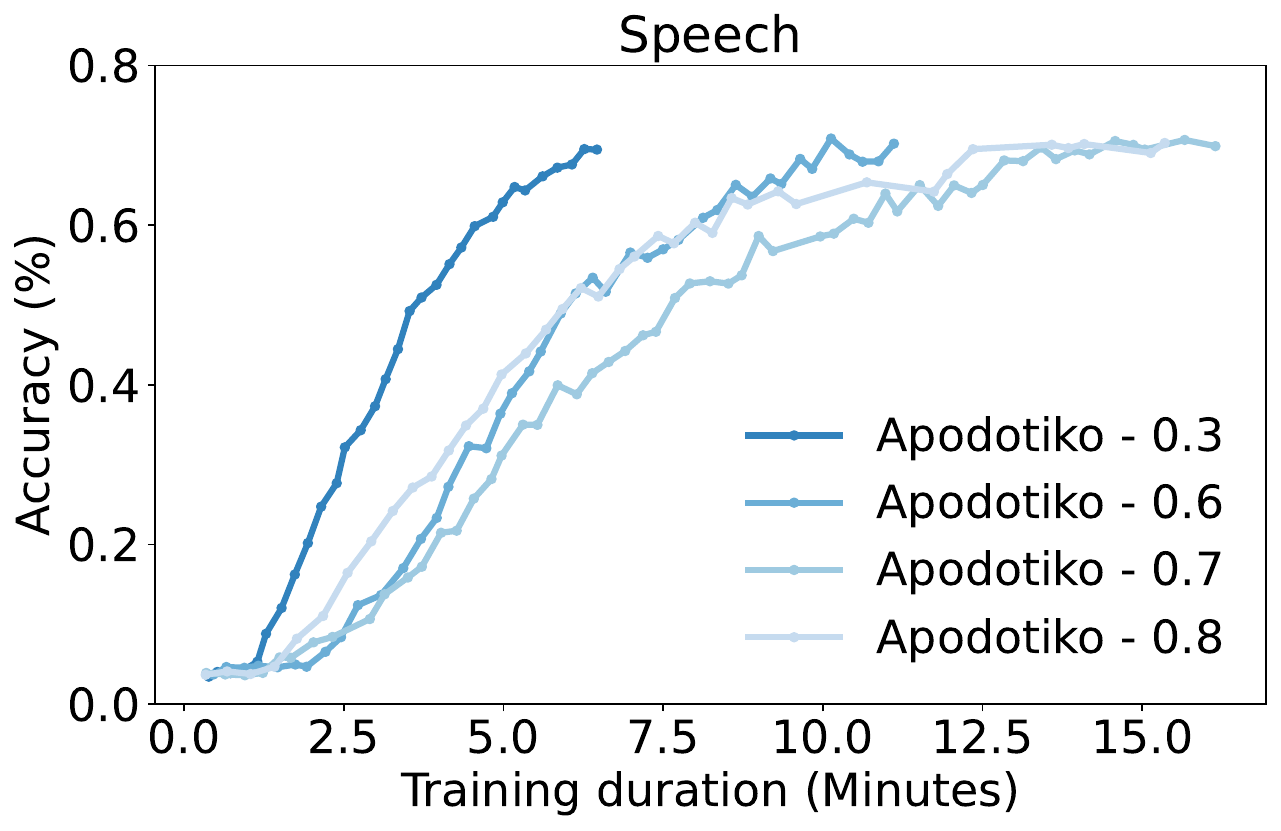}
        \caption{\emph{Google Speech}.}
        \label{fig:buffspeech}
\end{subfigure}
% \hspace{-5mm}
\vspace{-2mm}
\caption{Impact of different \emph{concurrencyRatios} (CRs).}
\label{fig:varyingbufferration}
%Better caption
% \shrinkspace
\end{figure}

Figure~\ref{fig:compcoldstartfinal} compares the cold start ratios among different FL strategies across multiple datasets. We observe that cold starts are significantly limited for the \emph{MNIST} and \emph{Google Speech} datasets due to the relatively brief round durations and the short intervals between each round and client invocation.  These factors contribute to a higher probability of a client being invoked again within a short timeframe, consequently minimizing the occurrence of cold starts. However, for the \emph{FEMNIST} and the \emph{Shakespeare} datasets, cold starts are more prominent due to larger average training round durations. In our experiments, we observe that \texttt{Apodotiko} consistently achieves a low cold start ratio across all datasets. This can be attributed to two reasons. First, 
the \emph{asynchronous aggregation} strategy used in \texttt{Apodotiko} triggers the invocation of the next round of clients as soon as a portion of the results becomes available. Second, our strategy incorporates a well-designed promotion mechanism (\S\ref{sec:scorebased}) that effectively prevents clients from missing multiple rounds over an extended period.

\begin{table}[t]
  \centering
  \begin{adjustbox}{width=0.8\columnwidth,  center}
  \begin{tabular}{|cc|c|c|c|c|}
    \hline
    \multicolumn{2}{|c|}{\diagbox{Strategy}{Dataset}}  & \multicolumn{1}{c|}{\textbf{MNIST (USD)}} & \multicolumn{1}{c|}{\textbf{FEMNIST (USD)}} & \multicolumn{1}{c|}{\textbf{Shakespeare (USD)}} & \multicolumn{1}{c|}{\textbf{Speech (USD)}}        \\ \hline
    \multicolumn{2}{|c|}{FedAvg~\cite{mcmahan2017communication}}                       & 1.13                                & \textbf{2.74}                         & 8.86                                      & 2.14                                        \\ \hline
    \multicolumn{2}{|c|}{FedProx~\cite{li2020federated}  }                    & 1.90                                & 3.83                                  & 10.63                                     & 2.37                                        \\ \hline
    \multicolumn{2}{|c|}{FedLesScan~\cite{elzohairy2022fedlesscan}}                   & \textbf{1.11}                       & 3.68                                  & 10.28                                     & \textbf{1.85}                               \\ \hline
    \multicolumn{2}{|c|}{SCAFFOLD~\cite{scaffold} }                    & 1.52                                & -                                     & 8.41                                      & -                                           \\ \hline
    \multicolumn{1}{|c|}{\multirow{4}{*}{Apodotiko}} & CR = 0.3                                 & 11.97                                 & 5.99                                      & \textbf{6.68}                        & 2.72 \\ \cline{2-6}
    \multicolumn{1}{|c|}{}                             & CR = 0.6                                 & 7.65                                  & 9.05                                      & 8.91                                 & 4.05 \\ \cline{2-6}
    \multicolumn{1}{|c|}{}                             & CR = 0.7                                 & 7.35                                  & 4.92                                      & 9.47                                 & 3.91 \\ \cline{2-6}
    \multicolumn{1}{|c|}{}                             & CR = 0.8                                 & 5.94                                  & 9.71                                      & 11.11                                & 3.14 \\ \hline
  \end{tabular}
  \end{adjustbox}
  \caption{Comparing total training cost (USD) across various FL strategies and datasets. The highlighted values represent the minimum costs for a particular dataset.}
  \label{tab:cost-comparison}
  % \shrinkspace
\end{table}

\vspace{-3mm}

\subsection{Comparing Time and Cost}
\label{sec:timecost}
Table~\ref {tab:duration-comparison} compares the total training time for the different FL strategies across multiple datasets. In our experiments, we observe that \texttt{Apodotiko} consistently outperforms other training strategies, particularly with a CR value of $0.3$. Table~\ref{tab:cost-comparison} compares the total training cost in USD for the different FL strategies and datasets.  Although \texttt{Apodotiko} isn't the most cost-effective strategy, it remains competitive with other methods. For instance, our strategy with a CR of 0.3 incurs a total training cost of $6.68$ USD for the \emph{Shakespeare} dataset and $2.72$ USD for the \emph{Google Speech} dataset. The increased costs can be attributed to the invocation of more clients, a consequence of the asynchronous nature of our strategy.

\vspace{-2mm}

\subsection{Sensitivity Analysis}
\label{sec:compbuffer}
We examine \texttt{Apodotiko}s' effectiveness across various environments and configurations. 

\textbf{Impact of different \emph{concurrencyRatios}}. Figures~\ref{fig:buffshakes} and~\ref{fig:buffspeech} show the effect of different CRs (\S\ref{sec:asyncaggregation}) on \texttt{Apodotiko} for the \emph{Shakespeare} and the \emph{Google Speech} datasets respectively. Our experiments on both datasets demonstrate that \texttt{Apodotiko} with a CR of 0.3 exhibits the fastest convergence rate compared to other concurrency ratios. For the \emph{Shakespeare} dataset, we observe a speedup of $1.34$x with a CR of $0.3$ compared to the ratio of $0.6$.  For the \emph{Speech} dataset, this speedup factor further increases to $1.7$x. This accelerated convergence is attributed to the controller's ability to trigger model aggregation with only 30 clients, significantly reducing the time between model updates. Additionally, our client selection algorithm (\S\ref{sec:selectingclients}) and stale weight aggregation function (\S\ref{sec:asyncaggregation}) effectively identify and select high-quality clients, ensuring that the aggregated model does not diverge. Table~\ref {tab:duration-comparison} and~\ref{tab:cost-comparison} also highlight the performance/cost of our strategy with different CR values against other FL approaches.

% Similarly, for the \emph{Speech} dataset, this improvement amounts to $1.7$x.

\begin{figure}[t]
\begin{subfigure}{0.15\textwidth}
    \centering
        \includegraphics[width=\columnwidth]{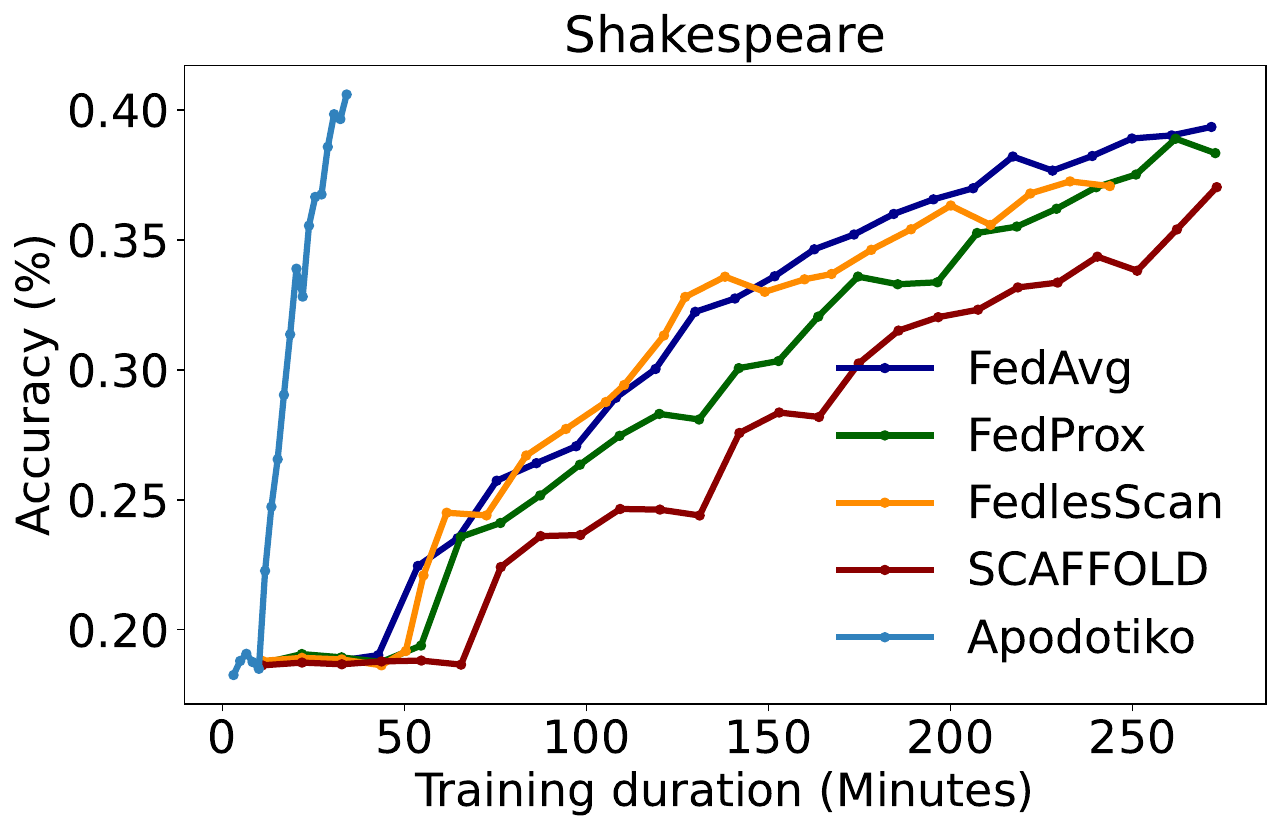}
        \caption{50/200 clients.}
        \label{fig:50clients}
\end{subfigure}
\begin{subfigure}{0.15\textwidth}
    \centering
        \includegraphics[width=\columnwidth]{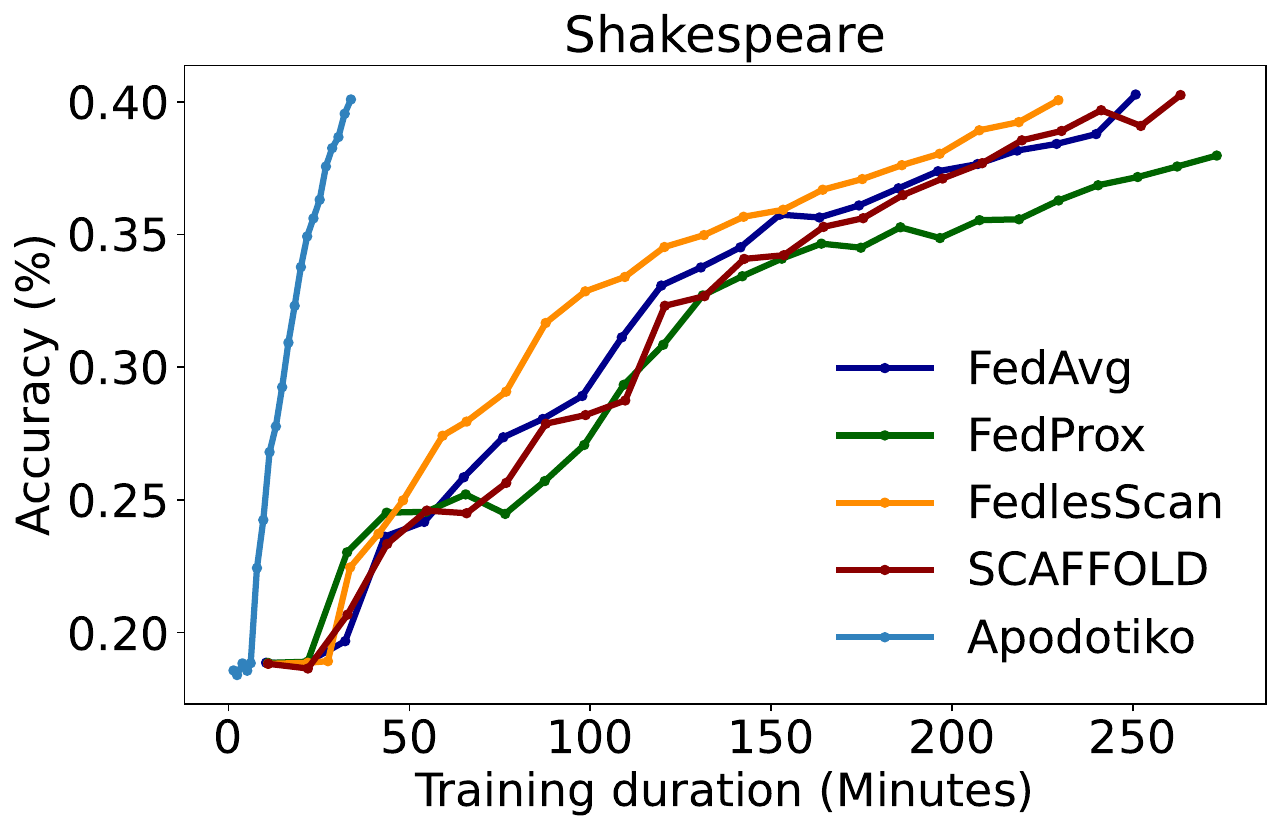}
        \caption{100/200 clients.}
        \label{fig:100clients}
\end{subfigure}
\begin{subfigure}{0.15\textwidth}
    \centering
        \includegraphics[width=\columnwidth]{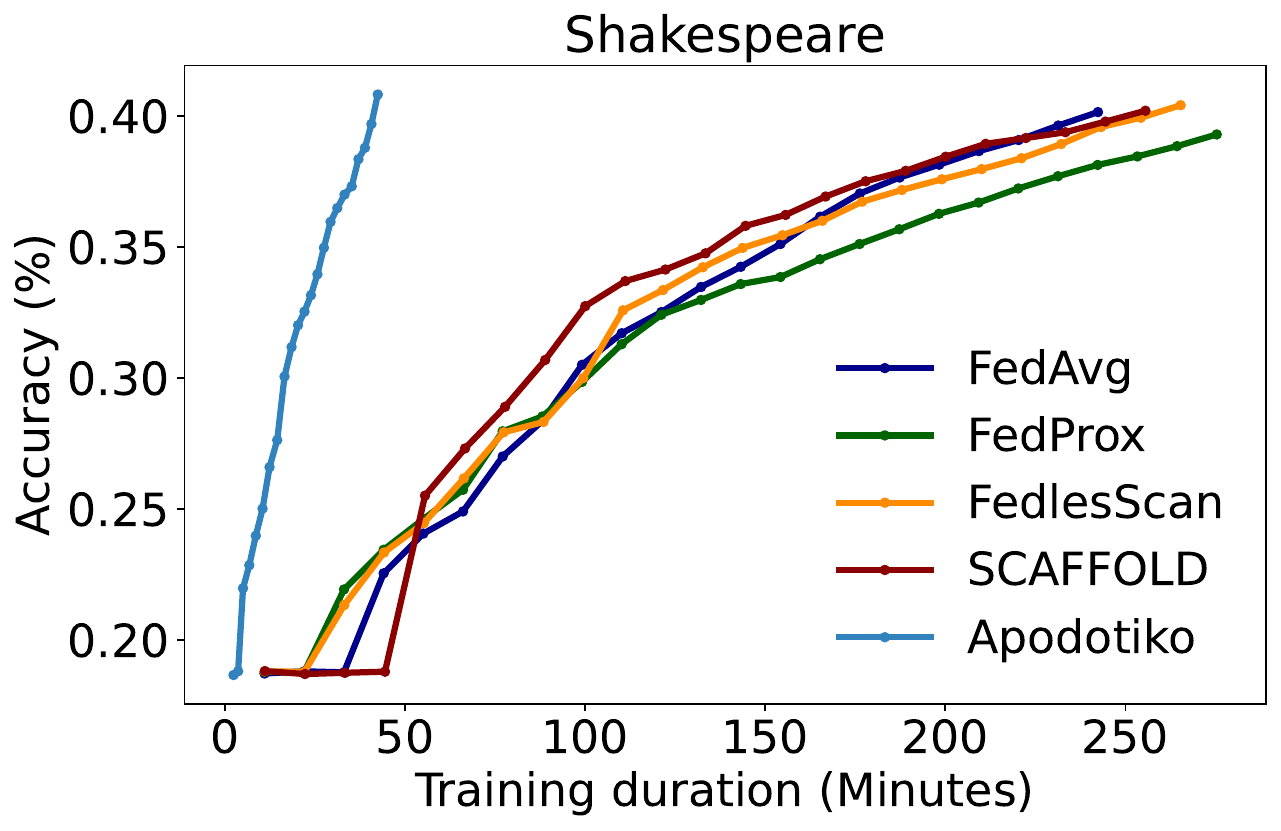}
        \caption{200/200 clients.}
        \label{fig:200clients}
\end{subfigure}
% \hspace{-5mm}
\vspace{-2mm}
\caption{Impact of different client sample sizes per round.}
\label{sec:figvaryingclients}
% \shrinkspace
%Better caption
\end{figure}

\textbf{Impact of different client sample sizes per round.}.
Figures~\ref{fig:50clients},~\ref{fig:100clients}, and~\ref{fig:200clients} show the performance of the different FL strategies with $50$, $100$, and $200$ clients selected for training in each round for the \emph{Shakespeare} dataset. Our experiments show that \texttt{FedAvg} and \texttt{FedProx} perform similarly regardless of the client sample size. Moreover, we observe \texttt{FedLesScan} performs best with $100$ clients per round.  This could be attributed to smaller cluster sizes, enhancing the probability of choosing sufficient clients from the same cluster for each training round. In contrast, we observe that \texttt{SCAFFOLD} performs better with increasing client sample size as the global variant~\cite{scaffold} becomes more accurate with more clients' results. Our observations about \texttt{SCAFFOLD} also align with the results presented in~\cite{li2022federated}. \texttt{Apodotiko} shows no significant impact on convergence rate with different sample sizes, as it only waits for a specific ratio of clients to complete the training before aggregating results and invoking new clients from the pool. We omit results for other datasets dues to space constraints but observe similar results.

% \vspace{-5mm}

\section{Conclusion and Future Work}
\label{sec:concfuture}
\vspace{-1mm}
In this paper, we presented \texttt{Apodotiko}, a novel \emph{asynchronous} scoring-based strategy that enables efficient serverless FL across clients with varying hardware resource configurations. We comprehensively evaluated our strategy against five other popular FL training approaches on multiple datasets. Our experiments highlight that \texttt{Apodotiko} converges faster and consistently minimizes cold starts in client function invocations. In the future, we plan to investigate the usage of an adaptive \emph{concurrencyRatio} based on the historical behavior of selected clients. This adaptive approach would potentially enable the inclusion of more results in the aggregation round by slightly delaying the model aggregation, thus preventing results from becoming stale.

% \vspace{-4mm}
% This approach would allow for the inclusion of more results in the aggregation round by delaying the model aggregation by a few seconds, thus preventing results from becoming stale.

% Furthermore, an adaptive buffer ratio could be introduced to adjust the waiting time based on the historical behavior of selected clients. This approach would allow for the inclusion of more results in the aggregation round by extending the waiting time by a few seconds, thus preventing results from becoming stale.

% We comprehensively evaluated our strategy against five other popular FL training strategies and demonstrated that \texttt{Apodotiko} converges faster, does introduce significant bias while selecting clients, and consistently minimizes cold starts client function cold starts. In the future, 

\section{Acknowledgements}
\vspace{-2mm}
The research leading to these results was funded by the German Federal Ministry of Education and Research (BMBF) in the scope of the Software Campus program under the grant agreement 01IS17049. 

\vspace{-4mm}
\bibliographystyle{IEEEtran}
\thispagestyle{empty}
\bibliography{parallelpgm}

% Generated by IEEEtran.bst, version: 1.14 (2015/08/26)
\begin{thebibliography}{10}
\providecommand{\url}[1]{#1}
\csname url@samestyle\endcsname
\providecommand{\newblock}{\relax}
\providecommand{\bibinfo}[2]{#2}
\providecommand{\BIBentrySTDinterwordspacing}{\spaceskip=0pt\relax}
\providecommand{\BIBentryALTinterwordstretchfactor}{4}
\providecommand{\BIBentryALTinterwordspacing}{\spaceskip=\fontdimen2\font plus
\BIBentryALTinterwordstretchfactor\fontdimen3\font minus \fontdimen4\font\relax}
\providecommand{\BIBforeignlanguage}[2]{{%
\expandafter\ifx\csname l@#1\endcsname\relax
\typeout{** WARNING: IEEEtran.bst: No hyphenation pattern has been}%
\typeout{** loaded for the language `#1'. Using the pattern for}%
\typeout{** the default language instead.}%
\else
\language=\csname l@#1\endcsname
\fi
#2}}
\providecommand{\BIBdecl}{\relax}
\BIBdecl

\bibitem{Privacy_and_big_data}
B.~M. Gaff \emph{et~al.}, ``Privacy and big data,'' \emph{Computer}, vol.~47, no.~6, pp. 7--9, 2014.

\bibitem{lecun2015deep}
Y.~LeCun, Y.~Bengio, and G.~Hinton, ``{Deep learning},'' \emph{nature}, vol. 521, no. 7553, pp. 436--444, 2015.

\bibitem{mcmahan2017communication}
B.~McMahan \emph{et~al.}, ``Communication-efficient learning of deep networks from decentralized data,'' in \emph{Artificial Intelligence and Statistics}.\hskip 1em plus 0.5em minus 0.4em\relax PMLR, 2017, pp. 1273--1282.

\bibitem{rieke2020future}
N.~Rieke \emph{et~al.}, ``The future of digital health with federated learning,'' \emph{NPJ digital medicine}, vol.~3, no.~1, p. 119, 2020.

\bibitem{huba2022papaya}
D.~Huba \emph{et~al.}, ``Papaya: Practical, private, and scalable federated learning,'' \emph{Proceedings of Machine Learning and Systems}, vol.~4, pp. 814--832, 2022.

\bibitem{serverlessfl}
\BIBentryALTinterwordspacing
M.~Chadha \emph{et~al.}, ``Towards federated learning using faas fabric,'' in \emph{Proceedings of the 2020 Sixth International Workshop on Serverless Computing}, ser. WoSC'20.\hskip 1em plus 0.5em minus 0.4em\relax New York, NY, USA: Association for Computing Machinery, 2020, p. 49–54. [Online]. Available: \url{https://doi.org/10.1145/3429880.3430100}
\BIBentrySTDinterwordspacing

\bibitem{fedless}
\BIBentryALTinterwordspacing
A.~Grafberger \emph{et~al.}, ``Fedless: Secure and scalable federated learning using serverless computing,'' in \emph{2021 IEEE International Conference on Big Data (Big Data)}, 2021, pp. 164--173. [Online]. Available: \url{https://doi.org/10.1109/BigData52589.2021.9672067}
\BIBentrySTDinterwordspacing

\bibitem{elzohairy2022fedlesscan}
\BIBentryALTinterwordspacing
M.~Elzohairy \emph{et~al.}, ``Fedlesscan: Mitigating stragglers in serverless federated learning,'' in \emph{2022 IEEE International Conference on Big Data (Big Data)}, 2022, pp. 1230--1237. [Online]. Available: \url{https://doi.org/10.1109/BigData55660.2022.10021037}
\BIBentrySTDinterwordspacing

\bibitem{jayaram2022lambda}
\BIBentryALTinterwordspacing
K.~Jayaram \emph{et~al.}, ``$\lambda$-fl: Serverless aggregation for federated learning,'' 2022. [Online]. Available: \url{http://tinyurl.com/3pkd2zks}
\BIBentrySTDinterwordspacing

\bibitem{jitfl}
K.~R. Jayaram \emph{et~al.}, ``Just-in-time aggregation for federated learning,'' in \emph{2022 30th International Symposium on Modeling, Analysis, and Simulation of Computer and Telecommunication Systems (MASCOTS)}, 2022, pp. 1--8.

\bibitem{jayaramadaptive}
K.~Jayaram \emph{et~al.}, ``Adaptive aggregation for federated learning,'' in \emph{2022 IEEE International Conference on Big Data (Big Data)}, 2022, pp. 180--185.

\bibitem{flox}
N.~Kotsehub \emph{et~al.}, ``Flox: Federated learning with faas at the edge,'' in \emph{2022 IEEE 18th International Conference on e-Science (e-Science)}, 2022, pp. 11--20.

\bibitem{fastgshare}
\BIBentryALTinterwordspacing
J.~Gu, Y.~Zhu, P.~Wang, M.~Chadha, and M.~Gerndt, ``Fast-gshare: Enabling efficient spatio-temporal gpu sharing in serverless computing for deep learning inference,'' in \emph{Proceedings of the 52nd International Conference on Parallel Processing}, ser. ICPP '23.\hskip 1em plus 0.5em minus 0.4em\relax New York, NY, USA: Association for Computing Machinery, 2023, p. 635–644. [Online]. Available: \url{https://doi.org/10.1145/3605573.3605638}
\BIBentrySTDinterwordspacing

\bibitem{chadha2024training}
\BIBentryALTinterwordspacing
M.~Chadha, P.~Khera, J.~Gu, O.~Abboud, and M.~Gerndt, ``Training heterogeneous client models using knowledge distillation in serverless federated learning,'' \emph{arXiv preprint arXiv:2402.07295}, 2024. [Online]. Available: \url{https://doi.org/10.48550/arXiv.2402.07295}
\BIBentrySTDinterwordspacing

\bibitem{fado}
\BIBentryALTinterwordspacing
C.~P. Smith, A.~Jindal, M.~Chadha, M.~Gerndt, and S.~Benedict, ``Fado: Faas functions and data orchestrator for multiple serverless edge-cloud clusters,'' in \emph{2022 IEEE 6th International Conference on Fog and Edge Computing (ICFEC)}, 2022, pp. 17--25. [Online]. Available: \url{https://doi.org/10.1109/ICFEC54809.2022.00010}
\BIBentrySTDinterwordspacing

\bibitem{fncapacitor}
\BIBentryALTinterwordspacing
A.~Jindal, M.~Chadha, S.~Benedict, and M.~Gerndt, ``Estimating the capacities of function-as-a-service functions,'' in \emph{Proceedings of the 14th IEEE/ACM International Conference on Utility and Cloud Computing Companion}, ser. UCC '21.\hskip 1em plus 0.5em minus 0.4em\relax New York, NY, USA: Association for Computing Machinery, 2022. [Online]. Available: \url{https://doi.org/10.1145/3492323.3495628}
\BIBentrySTDinterwordspacing

\bibitem{jindal2021function}
\BIBentryALTinterwordspacing
A.~Jindal, M.~Gerndt, M.~Chadha, V.~Podolskiy, and P.~Chen, ``Function delivery network: Extending serverless computing for heterogeneous platforms,'' \emph{Software: Practice and Experience}, vol.~51, no.~9, pp. 1936--1963, 2021. [Online]. Available: \url{https://onlinelibrary.wiley.com/doi/abs/10.1002/spe.2966}
\BIBentrySTDinterwordspacing

\bibitem{courier}
\BIBentryALTinterwordspacing
A.~Jindal, J.~Frielinghaus, M.~Chadha, and M.~Gerndt, ``Courier: Delivering serverless functions within heterogeneous faas deployments,'' in \emph{2021 IEEE/ACM 14th International Conference on Utility and Cloud Computing (UCC'21)}, ser. UCC '21.\hskip 1em plus 0.5em minus 0.4em\relax New York, NY, USA: Association for Computing Machinery, 2021. [Online]. Available: \url{https://doi.org/10.1145/3468737.3494097}
\BIBentrySTDinterwordspacing

\bibitem{postericdcs}
\BIBentryALTinterwordspacing
A.~Jindal, M.~Chadha, M.~Gerndt, J.~Frielinghaus, V.~Podolskiy, and P.~Chen, ``Poster: Function delivery network: Extending serverless to heterogeneous computing,'' in \emph{2021 IEEE 41st International Conference on Distributed Computing Systems (ICDCS)}, 2021, pp. 1128--1129. [Online]. Available: \url{https://doi.org/10.1109/ICDCS51616.2021.00120}
\BIBentrySTDinterwordspacing

\bibitem{chadha2021architecture}
\BIBentryALTinterwordspacing
M.~Chadha, A.~Jindal, and M.~Gerndt, ``Architecture-specific performance optimization of compute-intensive faas functions,'' in \emph{2021 IEEE 14th International Conference on Cloud Computing (CLOUD)}, 2021, pp. 478--483. [Online]. Available: \url{https://doi.org/10.1109/CLOUD53861.2021.00062}
\BIBentrySTDinterwordspacing

\bibitem{demystifying}
\BIBentryALTinterwordspacing
M.~Kiener, M.~Chadha, and M.~Gerndt, ``Towards demystifying intra-function parallelism in serverless computing,'' in \emph{Proceedings of the Seventh International Workshop on Serverless Computing (WoSC7) 2021}, ser. WoSC '21.\hskip 1em plus 0.5em minus 0.4em\relax New York, NY, USA: Association for Computing Machinery, 2021, p. 42–49. [Online]. Available: \url{https://doi.org/10.1145/3493651.3493672}
\BIBentrySTDinterwordspacing

\bibitem{openfaas}
\BIBentryALTinterwordspacing
OpenFaaS, ``{OpenFaaS - Serverless Functions Made Simple},'' 2019. [Online]. Available: \url{https://www.openfaas.com/ https://docs.openfaas.com/}
\BIBentrySTDinterwordspacing

\bibitem{gcloud-functions-2}
\BIBentryALTinterwordspacing
{Google Cloud}, ``{Cloud Functions Second Generation| Google Cloud},'' 2022. [Online]. Available: \url{http://tinyurl.com/3yapa4pv}
\BIBentrySTDinterwordspacing

\bibitem{castro2019rise}
P.~Castro \emph{et~al.}, ``The rise of serverless computing,'' \emph{Communications of the ACM}, vol.~62, no.~12, pp. 44--54, 2019.

\bibitem{hsieh2020non}
K.~Hsieh \emph{et~al.}, ``The non-iid data quagmire of decentralized machine learning,'' in \emph{International Conference on Machine Learning}.\hskip 1em plus 0.5em minus 0.4em\relax PMLR, 2020, pp. 4387--4398.

\bibitem{chai2020tifl}
Z.~Chai \emph{et~al.}, ``Tifl: A tier-based federated learning system,'' in \emph{Proceedings of the 29th international symposium on high-performance parallel and distributed computing}, 2020, pp. 125--136.

\bibitem{pisces}
Z.~Jiang \emph{et~al.}, ``Pisces: Efficient federated learning via guided asynchronous training,'' in \emph{Proceedings of the 13th Symposium on Cloud Computing}, ser. SoCC '22.\hskip 1em plus 0.5em minus 0.4em\relax New York, NY, USA: Association for Computing Machinery, 2022, p. 370–385.

\bibitem{fedprox}
A.~K. Sahu \emph{et~al.}, ``On the convergence of federated optimization in heterogeneous networks,'' \emph{arXiv preprint arXiv:1812.06127}, vol.~3, p.~3, 2018.

\bibitem{scaffold}
\BIBentryALTinterwordspacing
S.~P. Karimireddy \emph{et~al.}, ``Scaffold: Stochastic controlled averaging for on-device federated learning,'' \emph{CoRR}, vol. abs/1910.06378, 2019. [Online]. Available: \url{http://arxiv.org/abs/1910.06378}
\BIBentrySTDinterwordspacing

\bibitem{wu2020safa}
W.~Wu \emph{et~al.}, ``Safa: A semi-asynchronous protocol for fast federated learning with low overhead,'' \emph{IEEE Transactions on Computers}, vol.~70, no.~5, pp. 655--668, 2020.

\bibitem{fedbuff}
\BIBentryALTinterwordspacing
J.~Nguyen \emph{et~al.}, ``Federated learning with buffered asynchronous aggregation,'' \emph{CoRR}, vol. abs/2106.06639, 2021. [Online]. Available: \url{https://arxiv.org/abs/2106.06639}
\BIBentrySTDinterwordspacing

\bibitem{caldas2018leaf}
S.~Caldas \emph{et~al.}, ``{LEAF: A Benchmark for Federated Settings},'' in \emph{Workshop on Federated Learning for Data Privacy and Confidentiality, NeurIPS}, 2018, pp. 1--9.

\bibitem{osstalk}
``Fedless: Secure and scalable serverless federated learning,'' \url{https://osseu2023.sched.com/event/1OGe2}.

\bibitem{ray}
P.~Moritz \emph{et~al.}, ``Ray: A distributed framework for emerging {AI} applications,'' in \emph{13th USENIX Symposium on Operating Systems Design and Implementation (OSDI 18)}.\hskip 1em plus 0.5em minus 0.4em\relax Carlsbad, CA: USENIX Association, Oct. 2018, pp. 561--577.

\bibitem{Mothukuri2021}
V.~Mothukuri \emph{et~al.}, ``{A survey on security and privacy of federated learning},'' \emph{Future Generation Computer Systems}, vol. 115, pp. 619--640, feb 2021.

\bibitem{chard2020funcx}
\BIBentryALTinterwordspacing
R.~Chard \emph{et~al.}, ``Funcx: A federated function serving fabric for science,'' in \emph{Proceedings of the 29th International Symposium on High-Performance Parallel and Distributed Computing}, ser. HPDC '20.\hskip 1em plus 0.5em minus 0.4em\relax New York, NY, USA: Association for Computing Machinery, 2020, pp. 65--76. [Online]. Available: \url{https://doi.org/10.1145/3369583.3392683}
\BIBentrySTDinterwordspacing

\bibitem{li2020federated}
T.~Li \emph{et~al.}, ``Federated optimization in heterogeneous networks,'' 2020.

\bibitem{wang2020tackling}
J.~Wang \emph{et~al.}, ``Tackling the objective inconsistency problem in heterogeneous federated optimization,'' 2020.

\bibitem{cox2022aergia}
B.~Cox \emph{et~al.}, ``Aergia: leveraging heterogeneity in federated learning systems,'' in \emph{Proceedings of the 23rd ACM/IFIP International Middleware Conference}, 2022, pp. 107--120.

\bibitem{Oort}
\BIBentryALTinterwordspacing
F.~Lai \emph{et~al.}, ``Oort: Informed participant selection for scalable federated learning,'' \emph{CoRR}, vol. abs/2010.06081, 2020. [Online]. Available: \url{https://arxiv.org/abs/2010.06081}
\BIBentrySTDinterwordspacing

\bibitem{chai2021fedat}
Z.~Chai \emph{et~al.}, ``Fedat: a high-performance and communication-efficient federated learning system with asynchronous tiers,'' in \emph{Proceedings of the International Conference for High Performance Computing, Networking, Storage and Analysis}, 2021, pp. 1--16.

\bibitem{fedasync}
\BIBentryALTinterwordspacing
C.~Xie, S.~Koyejo, and I.~Gupta, ``Asynchronous federated optimization,'' \emph{CoRR}, vol. abs/1903.03934, 2019. [Online]. Available: \url{http://arxiv.org/abs/1903.03934}
\BIBentrySTDinterwordspacing

\bibitem{li2022federated}
Q.~Li \emph{et~al.}, ``Federated learning on non-iid data silos: An experimental study,'' in \emph{IEEE International Conference on Data Engineering}, 2022.

\bibitem{fedscale}
F.~Lai \emph{et~al.}, ``Fedscale: Benchmarking model and system performance of federated learning,'' in \emph{Proceedings of the First Workshop on Systems Challenges in Reliable and Secure Federated Learning}, 2021, pp. 1--3.

\bibitem{4paradigmk8svgpuscheduler}
\BIBentryALTinterwordspacing
S.~Li, Pei and Zheng, ``4paradigm/k8s-vgpu-scheduler: Open aios vgpu scheduler for kubernetes,'' 2022. [Online]. Available: \url{https://github.com/4paradigm/k8s-vgpu-scheduler}
\BIBentrySTDinterwordspacing

\bibitem{gcp.VM.pricing}
\BIBentryALTinterwordspacing
(2023) Pricing, compute engine: Virtual machines (vms), google cloud. [Online]. Available: \url{https://cloud.google.com/compute/all-pricing?authuser=2#gpus}
\BIBentrySTDinterwordspacing

\bibitem{gcp.function.pricing}
\BIBentryALTinterwordspacing
(2023) Pricing, cloud functions, google cloud. [Online]. Available: \url{https://cloud.google.com/functions/pricing}
\BIBentrySTDinterwordspacing

\end{thebibliography}

% \listoftodos

\end{document}